\documentclass[aip,prb,showpacs,twocolumn,10pt]{revtex4}
\usepackage{amsmath}
\usepackage{amsfonts}
\usepackage{amssymb}
\usepackage{graphicx}
\usepackage{appendix}
\providecommand{\U}[1]{\protect\rule{.1in}{.1in}}

\begin{document}
\title{Non-degenerate, three-wave mixing with the Josephson ring modulator}
\author{Baleegh Abdo}
\email{baleegh.abdo@yale.edu}
\author{Archana Kamal}
\author{Michel Devoret}
\affiliation{Department of Applied Physics, Yale University, New Haven, CT 06520, USA.}
\date{\today}

\begin{abstract}
The Josephson ring modulator (JRM) is a device, based on Josephson tunnel junctions, capable of performing non-degenerate mixing in the microwave regime without losses. The generic scattering matrix of the device is calculated by
solving coupled quantum Langevin equations. Its form shows that the device can
achieve quantum-limited noise performance both as an amplifier and a mixer.
Fundamental limitations on simultaneous optimization of performance metrics
like gain, bandwidth and dynamic range (including the effect of pump
depletion) are discussed. We also present three possible integrations of the
JRM as the active medium in a different electromagnetic environment. The
resulting circuits, named Josephson parametric converters (JPC), are discussed
in detail, and experimental data on their dynamic range are found to be in
good agreement with theoretical predictions. We also discuss future prospects
and requisite optimization of JPC as a preamplifier for qubit readout applications.
\end{abstract}

\maketitle


The photon energy of microwave radiation in the band from $4-8$ GHz ($\sim$ $8-4$ cm wavelength) is approximately $10^{5}$ smaller than that of the visible light. Yet, at a temperature $10^{4}$ smaller than room temperature, now routinely achievable with a dilution refrigerator, it is now possible to resolve the energy of single microwave photons \cite{DSnature}. There are three advantages of single photon microwave electronics when compared with quantum optics. First, signal shapes at carrier frequencies of a few GHz with a relative bandwidth of few percent can be controlled with much greater relative precision than their equivalent at a few hundreds of THz. This is partly due to the fact that microwave generators have more short term stability than lasers, but also because microwave components are
mechanically very stable, particularly when cooled, compared with traditional
optical components. Second, in single photon microwave electronics, the
on-chip circuitry can be well in the lumped element regime, and spatial mode
structure can be controlled more thoroughly and more reliably than in the
optical domain. Finally, there exists a simple, robust non-dissipative
component, the Josephson tunnel junction (JJ), whose non-linearity can be
ultra-strong even at the single photon level \cite{MDphysik}. Many quantum
signal processing functions have been realized using JJs, both digital and
analog, and this short review will not attempt to describe all of them. We
will focus on analog Josephson devices pumped with a microwave tone. They
recently led to microwave amplifiers working at the single photon level
\cite{CastellanosNat,JPCnature}. These novel devices have taken the work
pioneered by B. Yurke at Bell labs 25 years ago
\cite{YurkePRL,ParamYurkePRA,MovshovichPRL} to the point where actual
experiments can be performed using Josephson amplifiers as the first link in
the chain of measurement \cite{TeufelNatTech,QuantumJumps,QubitJPC}.
\par
In this paper, we address one particular subclass of analog signal processing
devices based on Josephson tunnel junction, namely those performing
non-degenerate three-wave mixing. Examples are Josephson circuits based on the
Josephson ring modulator \cite{JPCnaturePhys,Jamp} which we will describe
below. The Hamiltonian of such a device is of the form
\begin{eqnarray}
    H_{0}  & = & \frac{1}{2}\left( \frac{P_{X}^{2}}{\mathcal{M}_{X}}
    +\frac{P_{Y}^{2}}{\mathcal{M}_{Y}}
    +\frac{P_{Z}^{2}}{\mathcal{M}_{Z}}\right)
    \nonumber\\
    &  & + \frac{1}{2}\left(  \mathcal{K}_{X}X^{2}+\mathcal{K}_{Y}Y^{2} +\mathcal{K}_{Z}Z^{2}\right)  +KXYZ, \label{Hamiltonian1}%
\end{eqnarray}
where ($X$, $Y$, $Z$) and ($P_{X}$, $P_{Y}$, $P_{Z}$) are the generalized
position and momentum variables for the three independent oscillators,
$\mathcal{M}_{X,Y,Z}$ and $\mathcal{K}_{X,Y,Z}$ represent the ``mass" and
``spring constant" of the relevant oscillator (see table I), and $K$ is the
three-wave mixing constant which governs the non-linearity of the system. We
will discuss later how such simple minimal non-linear term can arise. The
classical equation of motions for the standing waves in such a device are
symmetric and are given by:
\begin{eqnarray}
    \overset{\cdot\cdot}{X}+\gamma_{a}\overset{\cdot}{X}+\omega_{a}^{2} X+K^{\prime}YZ  &  =x\left(  t\right)  \cos\omega_{a}t,
    \label{non-degenerate1}\\
    \overset{\cdot\cdot}{Y}+\gamma_{b}\overset{\cdot}{Y}+\omega_{b}^{2} Y+K^{\prime}XZ  &  =y\left(  t\right)  \cos\omega_{b}t,
    \label{non-degenerate2}\\
    \overset{\cdot\cdot}{Z}+\gamma_{c}\overset{\cdot}{Z}+\omega_{c}^{2}Z+K^{\prime}XY  &  =z\left(  t\right)  \cos\omega_{c}t,
    \label{non-degenerate3}%
\end{eqnarray}
where $K^{\prime}=K/\mathcal{M}$ (we assume, for simplicity, equal masses $\mathcal{M}_{X,Y,Z}=\mathcal{M}$) and $\omega_{a,b,c}$ $=$ $\sqrt
{\mathcal{K}_{X,Y,Z}/\mathcal{M}}$ are the angular resonant frequencies of the three coordinates satisfying
\begin{equation}
    \omega_{a}<\omega_{b}\,<\omega_{c}=\omega_{a}+\omega_{b}.
\end{equation}
\begin{table*}[t!]
\centering
\begin{tabular}[c]{|c|c|c|c|}\hline
    $X $ \; \textrm{(position)} & $P$ \; \textrm{(momentum)} & $\mathcal{M}$
    \;\textrm{(mass)} & $\mathcal{K}$ \; \textrm{(spring constant)}
    \\
    \hline\hline
    $\Phi$ \; \textrm{(flux) } & $Q$ & $C$ \; \textrm{(capacitance)} & $L^{-1}
    $\\
    \hline
    $Q$ \; \textrm{(charge) } & $\Phi$ & $L$ \; \textrm{(inductance)} & $C^{-1}
    $\\
    \hline
\end{tabular}
\label{Table1}
\caption{Generalized variables and parameters for the system of
oscillators described by Eq. (1). The variables and parameters listed in the
first line apply to the case of a mechanical oscillator, whereas the ones
listed in the second and third lines are adapted for describing an LC
oscillator with parallel and series dissipations, respectively, as shown in
Fig. 1 and Fig. 16 (top panel) for the parallel case, and Fig. 16 (bottom
panel) for the series case. }%
\end{table*}
We also suppose the oscillators are well in the underdamped regime
\begin{align}
    \gamma_{a}  &  \ll\omega_{a},\\
    \gamma_{b}  &  \ll\omega_{b},\\
    \gamma_{c}  &  \ll\omega_{c},
\end{align}
a sufficient but not strictly necessary hypothesis, which has the principal
merit of keeping the problem analytically soluble under the conditions of
interest. It is worth noting that the system is non-degenerate both spatially
and temporally. On the other hand, it is important to suppose that the
envelope functions $x(t)$, $y\left(  t\right)  $ and $z\left(  t\right)  $ of
the drive signals are supposed to be slow compared to the respective drive
frequencies $\omega_{b}-\omega_{a}\gg\gamma_{a}+\gamma_{b}$.
\par
The equations (\ref{non-degenerate1}-\ref{non-degenerate3}) must be contrasted
with that of a degenerate three-wave mixing device for which two cases are
possible. In the first case, where the $Y$ and $Z$ degrees of freedom have
merged into a single oscillator, the Hamiltonian has a non-linear term of the
form $KXZ^{2}$ and the equations read:
\begin{align}
    \overset{\cdot\cdot}{X}+\gamma_{a}\overset{\cdot}{X}+\omega_{a}^{2}X+K^{\prime}Z^{2}  &  =x\left(  t\right)  \cos\omega_{a}t,\\
    \overset{\cdot\cdot}{Z}+\gamma_{c}\overset{\cdot}{Z}+\omega_{c}^{2}Z+2K^{\prime}ZX  &  =z\left(  t\right)  \cos\omega_{c}t.
\end{align}
This is the case of electromechanical resonators \cite{TeufelNature} in which
one of the capacitance plates of a microwave oscillator ($Z$) is itself the
mass of a mechanical resonator ($X$). There $\omega_{c}\gg\omega_{a}$, and
pumping the microwave oscillator in the vicinity of $\omega_{c}-\omega_{a}$
leads to cooling of the mechanical oscillator provided $\gamma_{c}\gg
\gamma_{a}$. In the second case, it is the $X$ and the $Y$ degrees of freedom
that merge into a single oscillator, leading to a non-linear term in the
Hamiltonian of the form $KX^{2}Z$. The equations then read
\begin{align}
    \overset{\cdot\cdot}{X}+\gamma_{a}\overset{\cdot}{X}+\omega_{a}^{2}X+2K^{\prime}XZ  &  =x\left(  t\right)  \cos\omega_{a}t,
    \label{degenerate 2}\\
    \overset{\cdot\cdot}{Z}+\gamma_{c}\overset{\cdot}{Z}+\omega_{c}^{2}Z+K^{\prime}X^{2}  &  =z\left(  t\right)  \cos\omega_{c}t 
    \label{degenerate 3}%
\end{align}
and we have now
\begin{equation}
    \omega_{c}=2\omega_{a}. 
    \label{degenerate 1}%
\end{equation}
This case is implemented in Josephson circuits as a dcSQUID whose flux is
driven by a microwave oscillating signal at twice the plasma frequency of the
SQUID \cite{Yamamoto}. When $z\left(  t\right)  =z_{d}\gg K^{\prime}X^{2}$
(so-called \textquotedblleft stiff" or \textquotedblleft non-depleted" pump
condition), the system of equations (\ref{degenerate 2},\ref{degenerate 3})
reduces to the parametrically driven oscillator equation
\begin{equation}
    \overset{\cdot\cdot}{X}+\gamma_{a}\overset{\cdot}{X}+\omega_{a}^{2}
    \left[1+\frac{K^{\prime}z_{d}}{\gamma_{c}\omega_{c}}\sin\left(  \omega_{c}t\right)\right]  X=x\left(  t\right)  \cos\omega_{a}t. \label{parametric oscillator}%
\end{equation}
Note that there is, in addition to the parametric drive on the left hand side,
a small perturbing drive signal $x\left(  t\right)  \cos\omega_{a}t$ on the
right hand side. The theory of the degenerate parametric amplifier starts with
this latter equation, the term $\frac{K^{\prime}z_{d}}{\gamma_{c}\omega_{c}%
}\sin\left(  \omega_{c}t\right)  $ corresponding to the pump and $x\left(
t\right)  \cos\omega_{a}t$ corresponding to the input signal. The output
signal is obtained from a combination of the loss term $\gamma_{a}\dot{X}$ and
the input signal.
\par
In the context of Josephson devices, another route to the effective parametric
oscillator of equation (\ref{parametric oscillator}) can be obtained by a
driven, Duffing-type oscillator \cite{VijayJBAreview,JBA}. This system
(Josephson bifurcation amplifier) has only one spatial mode and quartic
non-linearity,
\begin{equation}
\overset{\cdot\cdot}{X}+\gamma_{a}\overset{\cdot}{X}+\omega_{a}^{2}X-\lambda
X^{3}=\left[  z_{d}+x\left(  t\right)  \right]  \cos\omega_{d}t.
\label{parametric oscillator2}%
\end{equation}
Driven by a strong tone $z_{d}\cos\omega_{d}t$ in the vicinity of the
bifurcation occurring at
\begin{align}
    \omega_{d}  &  =\omega_{a}-\frac{\sqrt{3}}{2}\gamma_{a},\\
    z_{d}  &  =\frac{128}{27}\sqrt{\frac{\gamma_{a}^{3}\omega_{a}}{3\lambda}},
\end{align}
it will lead to an equation of the form (\ref{parametric oscillator}) for
small deviations around the steady-state solution. It will, therefore, amplify
the small drive modulation signal $x\left(  t\right)  $ of equation
(\ref{parametric oscillator2}) [\onlinecite{Hatridge}]. Similar amplifying effects can
be found in pumped superconducting microwave resonators without Josephson
junctions \cite{ImBaleegh,TholenAPL,Zmuidzinas}.
\par
In the following section, we will treat Eqs. (\ref{non-degenerate1}%
-\ref{non-degenerate3}) using input-output theory \cite{LinearScatteringNote}
and obtain the quantum-mechanical scattering matrix of the signal and idler
amplitudes in the stiff-pump approximation. This allows us to find the photon
gain of the device in its photon amplifier mode as a function of the pump
amplitude, and the corresponding reduction of bandwidth. We then discuss the
implementation of the device using a ring of four Josephson junctions
flux-biased at half-quantum in Sec. II. It is the non-dissipative analogue of
the semiconductor diode ring modulator \cite{Pozar}. In Sec. III, we treat the
finite amplitude of signals and establish useful relations between the dynamic
range, gain and bandwidth. In Sec. IV we introduce the Josephson parametric
converter (JPC) as an example of a non-degenerate, three-wave mixing device
operating at the quantum limit. We present three different realizations
schemes for the JPC and point out their practical advantages and limitations.
In Sec. V we present experimental results for different JPC\ devices and
compare the data with the maximum bounds predicted by theory. We follow this
with a discussion, in Sec. VI, of general requirements for an amplifier to
meet the needs of qubit readout and how the maximum input power of the device
can be increased by two orders of magnitude beyond typical values achieved
nowadays. We conclude with a brief summary of our results in Sec. VII.
%
%
\section{Input-output treatment of a generic non-degenerate, three-wave mixing device}
\begin{figure}[h]
\begin{center}
\includegraphics[width=\columnwidth]{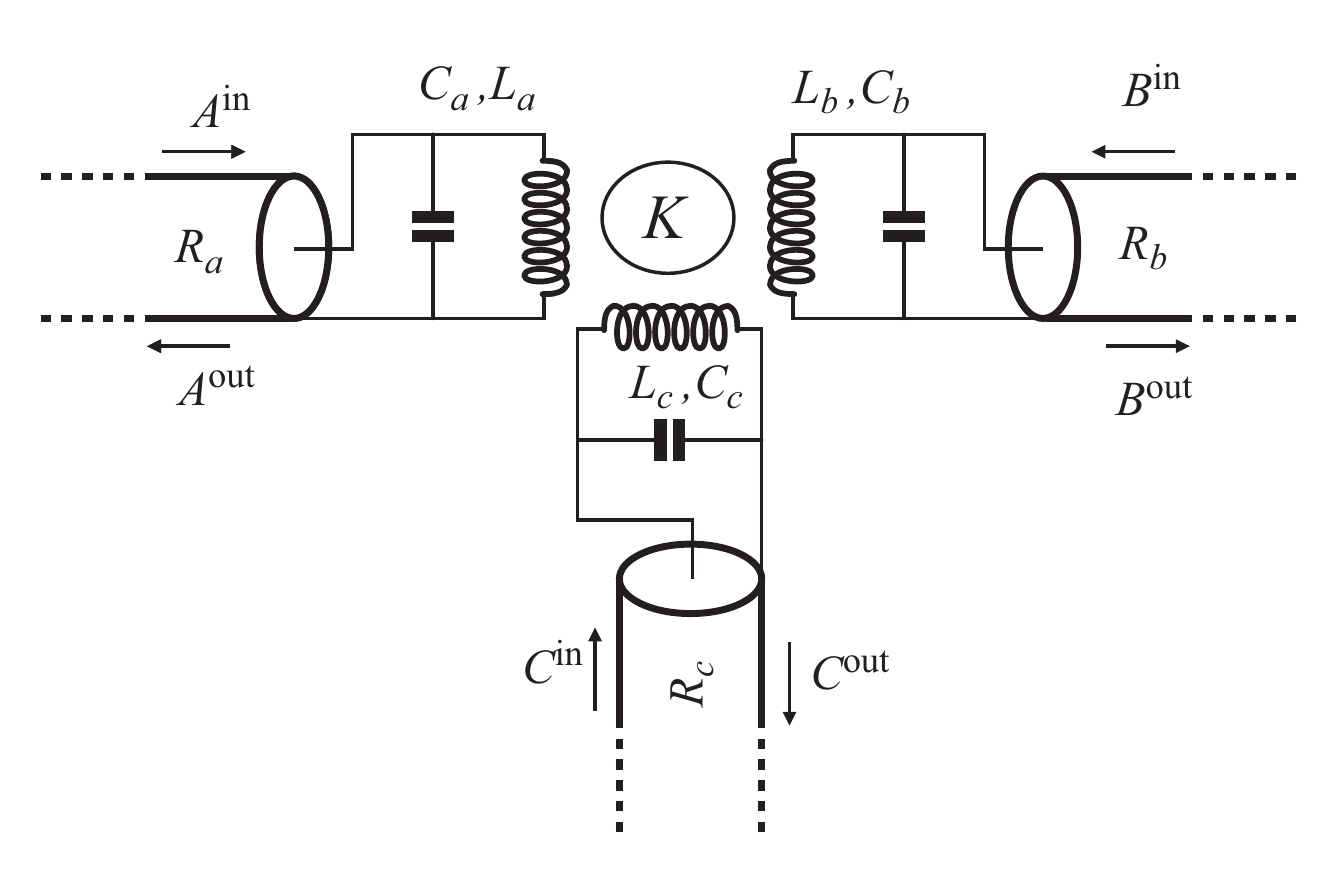}\\
\caption{General non-degenerate three-wave mixing device consisting of three
LC oscillators coupled by a non-linear medium, giving a trilinear term in the
Hamiltonian of the form $K\Phi_{a}\Phi_{b}\Phi_{c}$ where the fluxes
$\Phi_{a,b,c}$ are those of the inductors. Each oscillator is fed by a
transmission line with characteristic impedance $R_{a,b,c}$.}%
\label{three_osc_fig}%
\end{center}
\end{figure}

%
%
The three oscillators of Eqs. (\ref{non-degenerate1}-\ref{non-degenerate3})
correspond to three quantum LC oscillators coupled by a non-linear, trilinear
mutual inductance, whose mechanism we will discuss in the next section. They
are fed by transmission lines which carry excitations both into and out of the
oscillators, as shown on Fig. \ref{three_osc_fig}. The Hamiltonian of the
system is (leaving out the transmission lines for the moment),%
\begin{align}
    \frac{H_{0}}{\hbar}  &  =\omega_{a}a^{\dag}a+\omega_{b}b^{\dag}b+\omega_{c}c^{\dag}c\nonumber\\
    &  +g_{3}\left(  a+a^{\dag}\right)  \left(  b+b^{\dag}\right)  \left(c+c^{\dag}\right)  ,
\end{align}
where $a$, $b$ and $c$ are the annihilation operators associated with each of
the three degrees of freedom. Their associated angular frequencies are given
in terms of the inductances and capacitances as
\begin{equation}
    \omega_{a,b,c}=\frac{1}{\sqrt{L_{a,b,c}C_{a,b,c}}}.
\end{equation}
The bosonic operators of different modes (a, b, c) commute with each other and
those associated with the same mode satisfy the usual commutation relations of
the form
\begin{equation}
    \left[  a,a^{\dag}\right]  =1.
\end{equation}
The link between the mode amplitude such as $X$, which represents the flux
through the inductance of the oscillator, and a quantum operator such as $a$
can be written as,
\begin{equation}
    X=X^{ZPF}\left(  a+a^{\dag}\right),
\end{equation}
where ``ZPF" stands for ``zero-point fluctuations" and
\begin{align}
    X^{ZPF}  &  =\sqrt{\frac{\hbar Z_{a}}{2}},\\
    Z_{a}  &  =\sqrt{\frac{L_{a}}{C_{a}}},
\end{align}
the last equation defining the impedance of the oscillator, equal to the
modulus of the impedance on resonance of either the inductance or the
capacitance. The link between $K$ and $g_{3}$ is therefore
\begin{equation}
    \hbar g_{3}=KX^{ZPF}Y^{ZPF}Z^{ZPF}.
\end{equation}
We now work in the framework of Rotating Wave Approximation (RWA), in which we
only keep terms commuting with the total photon number
\begin{equation}
    \frac{H_{0}^{\mathrm{RWA}}}{\hbar}
    =\omega_{a}a^{\dag}a+\omega_{b}b^{\dag}b+\omega_{c}c^{\dag}c+g_{3}\left(  a^{\dag}b^{\dag}c+abc^{\dag}\right).
\end{equation}
Treating in RWA the coupling of each oscillator with a transmission line
carrying waves in and out of the oscillator (see Appendix for complements of
the next 6 equations), one arrives at three coupled quantum Langevin equations
for $a\left(t\right)$, $b\left(t\right)$ and $c\left( t\right)$:
\begin{align}
    \frac{\mathrm{d}}{\mathrm{d}t}a  
    & = -i\omega_{a}a-ig_{3}b^{\dag}c-\frac{\gamma_{a}}{2}a+\sqrt{\gamma_{a}}\tilde{a}^{\mathrm{in}}\left(t\right),
    \nonumber\\
    \frac{\mathrm{d}}{\mathrm{d}t}b  
    & = -i\omega_{b}b-ig_{3}a^{\dag}c-\frac{\gamma_{b}}{2}b+\sqrt{\gamma_{b}}\tilde{b}^{\mathrm{in}}\left(t\right),
    \nonumber\\
    \frac{\mathrm{d}}{\mathrm{d}t}c  &  =-i\omega_{c}c-ig_{3}ab-\frac{\gamma_{c}}{2}c+\sqrt{\gamma_{c}}\tilde{c}^{\mathrm{in}}\left(t\right),
\label{threeAmpEqs}
\end{align}
In these equations, the second term in the right hand side corresponds to the
non-linear term producing photon conversion. The third term says that photons
introduced in one resonator leave with a rate
\begin{equation}
    \gamma_{a,b,c}=\omega_{a,b,c}\frac{Z_{a,b,c}}{R_{a,b,c}},
\end{equation}
with the resistances $R_{a,b,c}$ denoting the characteristic impedances of the
transmission lines. Finally, in the fourth term of the Langevin equations, the
input fields such as $\tilde{a}^{\mathrm{in}}\left(t\right)$ correspond to
the negative frequency component of the drive terms in the classical
equations. They obey the relation
\begin{equation}
    \tilde{a}^{\mathrm{in}}(t)=\frac{1}{\sqrt{2\pi}}\int_{0}^{+\infty}a^{\mathrm{in}}\left[  \omega\right]  e^{-i\omega t}\mathrm{d}\omega,
\end{equation}
where $a^{\mathrm{in}}[\omega]$ are the usual field operators obeying the commutation relations
\begin{equation}
    \left[a^{\mathrm{in}}\left[  \omega\right]  ,a^{\mathrm{in}}\left[\omega^{\prime}\right]  \right]  
    =\mathrm{sgn}\left(\frac{\omega-\omega^{\prime}}{2}\right)  \delta\left(\omega+\omega^{\prime}\right)
\end{equation}
in which $\omega$ denotes a frequency that can be either positive or negative.
The transmission lines thus both damp and drive the oscillators. The incoming
field operator treats the drive signals and the Nyquist equilibrium noise of
the reservoir on the same footing. Photon spectral densities $\mathcal{N}^{\mathrm{in}}[\omega]$ of the incoming fields, introduced by relations of the form
\begin{equation}
    \left\langle \left\{a^{\mathrm{in}}\left[\omega\right], a^{\mathrm{in}}\left[\omega^{\prime}\right]\right\}\right\rangle 
    =2\mathcal{N}_{a}^{\mathrm{in}}\left[  \frac{\omega-\omega^{\prime}}{2}\right]\delta\left(\omega+\omega^{\prime}\right), 
    \label{Na_in_first}
\end{equation}
have the value
\begin{align}
    \mathcal{N}_{a}^{\mathrm{in}}\left[  \omega\right]   
    & = \frac{\mathrm{sgn}\left(  \omega\right)  }{2}\coth\left(  \frac{\hbar\omega}{2k_{B}T}\right)
    \nonumber\\
    &  + 2\pi P_{a}^{\mathrm{in}}\left[  \delta\left(  \omega-\omega_{1}\right)+\delta\left(  \omega+\omega_{1}\right)  \right], 
    \label{Na_in_sec}%
\end{align}
where $P_{a}^{\mathrm{in}}$ is the photon flux of the incoming drive signal at
angular frequency $\omega_{1}$ (in units of photons per unit time) and $T$ is
the temperature of the electromagnetic excitations of the line. Note that the
dimensionless function $\mathcal{N}_{a}^{\mathrm{in}}\left[  \omega\right]  $
is defined for both positive and negative frequencies. It is symmetric
$\mathcal{N}_{a}^{\mathrm{in}}\left[  \omega\right]  =\mathcal{N}%
_{a}^{\mathrm{in}}\left[  -\omega\right]  $ and its value at frequency
$\left\vert \omega\right\vert $ represents the average number of photons per
unit time per unit bandwidth in the incoming signal, which in the high
temperature limit is $k_{B}T/\left(  \hbar\left\vert \omega\right\vert
\right)  $. It includes the $\frac{1}{2}$ contribution of zero-point quantum noise.
\par
\begin{figure}
[h]
\begin{center}
\includegraphics[width=\columnwidth]{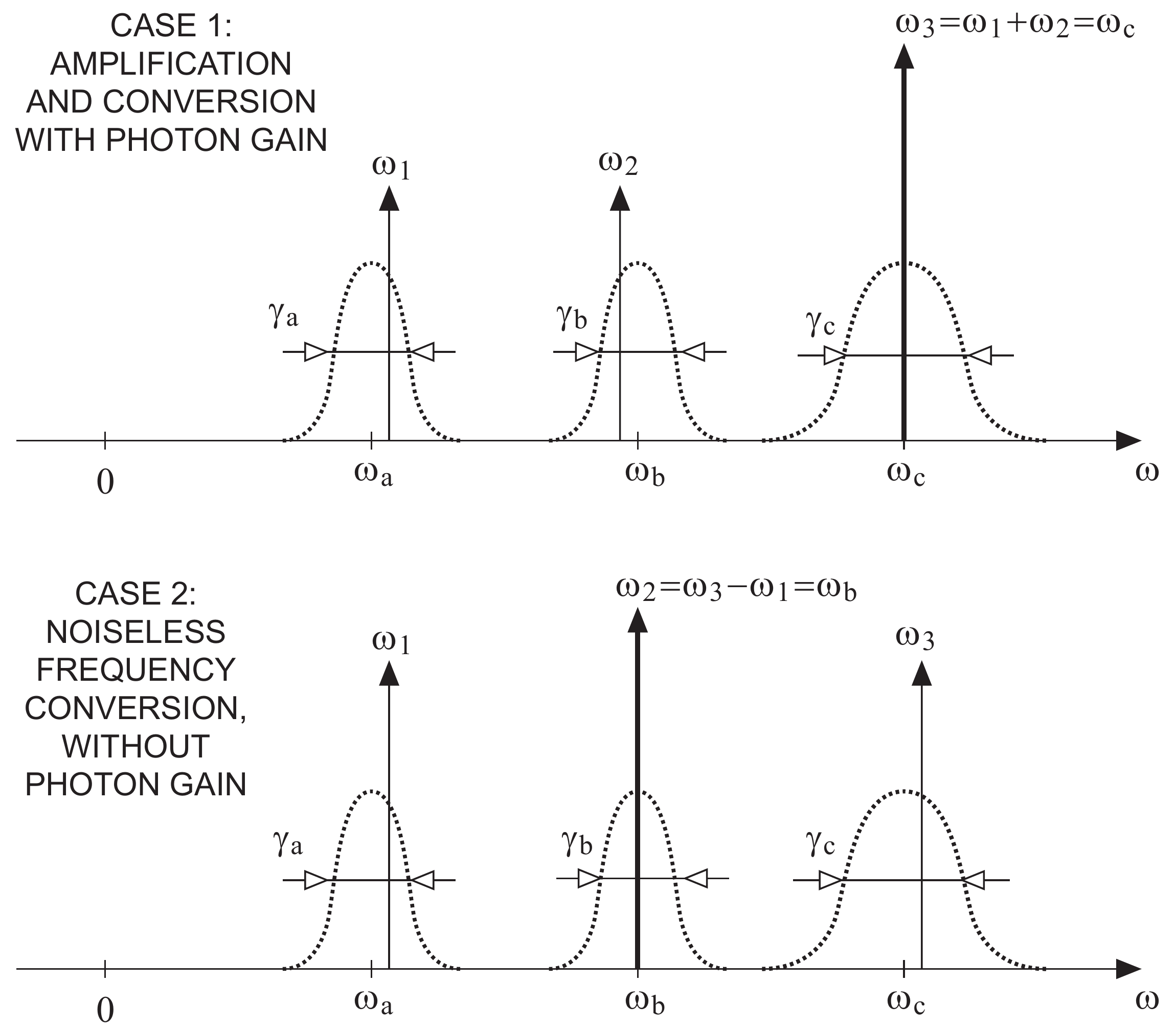}\\
\caption{Characteristic frequency landscape of non-degenerate three-wave
mixing devices. Three separate oscillators have resonant frequencies
$\omega_{a}<\omega_{b}<\omega_{c}=\omega_{a}+\omega_{b}$. They are fed by
transmission lines, giving them a full linewidth at half-maximum $\gamma_{a}$,
$\gamma_{b}$ and $\gamma_{c}$ respectively. The non-linear coupling strength,
expressed in photon amplitude language, is much smaller than these linewidths.
The device can be pumped at $\omega_{c}$ and operates then as a
phase-preserving amplifier with photon gain for frequencies $\omega_{a}$ and
$\omega_{b}$ (top), or it can be pumped at one of the two lower frequencies
$\omega_{a}$ or $\omega_{b}$ and operates then as a noiseless frequency
converter or dynamical cooler, upconverting signals into oscillator at
$\omega_{c}$ (bottom). In this figure, the spectral density of weak signal
corresponds to thin arrows whereas the spectral densities of pump signals
corresponds to thick arrows.}%
\label{frequencies}%
\end{center}
\end{figure}

It is worth insisting that we treat the non-linear coupling strength as a
perturbation compared with the influence of the reservoirs, treated themselves
as a perturbation compared with the Hamiltonian of the oscillators:
\begin{equation}
    g_{3}\ll\gamma_{a},\gamma_{b}<\gamma_{c}\ll\omega_{a},\omega_{b}<\omega_{c}=\omega_{a}+\omega_{b}.
\end{equation}
In general, only one strong drive tone is applied to one of the resonators and
is called the ``pump". Two cases must then be distinguished at this stage, as
shown in Fig. \ref{frequencies}:
\par
Case 1 (amplification and frequency conversion with photon gain): the pump
tone is applied to the $c$ resonator. The device is usually used as an
amplifier \cite{JPCnature,Jamp}. It can also be used as a two-mode squeezer
\cite{JPCSqueezer}.
\par
Case 2 (noiseless frequency conversion without photon gain): the pump tone is
applied to either the $a$ or $b$ resonator \cite{BSconv}. The device is useful
as a noiseless up- and down-converter and can perform dynamical cooling of the
lowest energy oscillator, transferring its spurious excitations to the highest
frequency one, which is more easily void of any excitations and plays the role
of a cold source.
%
%
\subsection{Photon gain (case 1)}
\begin{figure}
[h]
\begin{center}
\includegraphics[width=\columnwidth]{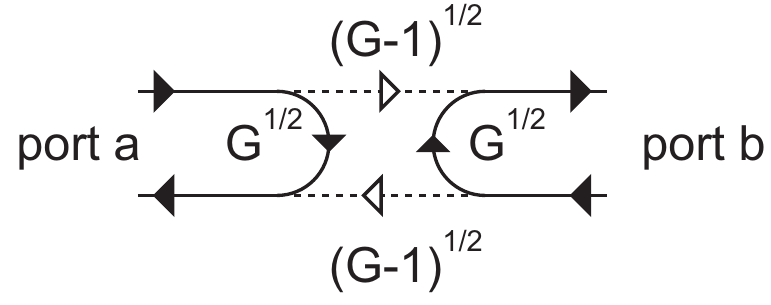}\\
\caption{An amplifier reaching the quantum limit must have a minimal
scattering matrix, with the signal in port $a$ being reflected with amplitude
gain $G^{1/2}$ while the signal in port $b$ is phase-conjugated and
transmitted to port $a$ with amplitude gain $(G-1)^{1/2}$. This can be
realized in case 1 of Fig. \ref{frequencies}.}%
\label{minimal-amp}%
\end{center}
\end{figure}

%
%
We will first suppose that the pump is ``stiff", namely
\begin{align}
\left\vert \left\langle \tilde{c}^{\mathrm{in}}\right\rangle \right\vert ^{2}
&  \gg1\\
\gamma_{c}  &  \gg\gamma_{a},\gamma_{b}%
\end{align}
This means that the pump tone will not be easily depleted despite the fact
that its photons are converted into the signal and idler photons at
$\omega_{a}$ and $\omega_{b}$. For solving the quantum Langevin equations, we
replace the pumped oscillator annihilation operator $c$ by its average value
in the coherent state produced by the pump as
\begin{equation}
c\left(  t\right)  \rightarrow\left\langle c\left(  t\right)  \right\rangle
=\sqrt{\bar{n}_{c}}e^{-i\left(  \omega_{c}t+\phi\right)  }.
\end{equation}
The Langevin equations can then be transformed into the linear equations (see
equation (\ref{IOT}) of Appendix)
\begin{multline}
\left[
\begin{array}
[c]{cc}%
O_{a}^{+} & ig_{b}^{a}e^{-i\omega_{c}t}\\
-ig_{a}^{b\ast}e^{+i\omega_{c}t} & O_{b}^{+\ast}%
\end{array}
\right]  \left[
\begin{array}
[c]{c}%
\tilde{a}^{\mathrm{out}}\\
\tilde{b}^{\mathrm{out}\dagger}%
\end{array}
\right]  =\\
-\left[
\begin{array}
[c]{cc}%
O_{a}^{-} & ig_{b}^{a}e^{-i\omega_{c}t}\\
-ig_{a}^{b\ast}e^{+i\omega_{c}t} & O_{b}^{-\ast}%
\end{array}
\right]  \left[
\begin{array}
[c]{c}%
\tilde{a}^{\mathrm{in}}\\
\tilde{b}^{\mathrm{in}\dagger}%
\end{array}
\right]  ,
\end{multline}
where
\begin{align}
O_{a,b}^{\pm}  &  =\frac{\mathrm{d}}{\mathrm{d}t}+i(\omega_{a,b}\mp
i\Gamma_{a,b}),\\
\Gamma_{a,b}  &  =\frac{\gamma_{a,b}}{2},\\
g_{b,a}^{a,b}  &  =g_{3}\sqrt{\bar{n}_{c}}e^{-i\phi}\sqrt{\frac{\Gamma_{a,b}%
}{\Gamma_{b,a}}}.
\end{align}
After a Fourier transform, we obtain in the frequency domain, a simpler
relation
\begin{multline}
\left[
\begin{array}
[c]{cc}%
h_{a}\left[  \omega_{1}\right]  & +ig_{b}^{a}\\
-ig_{a}^{b\ast} & h_{b}^{\ast}\left[  \omega_{2}\right]
\end{array}
\right]  \left[
\begin{array}
[c]{c}%
a^{\mathrm{out}}\left[  +\omega_{1}\right] \\
b^{\mathrm{out}}\left[  -\omega_{2}\right]
\end{array}
\right]  =\label{in-out-amp5}\\
\left[
\begin{array}
[c]{cc}%
h_{a}^{\ast}\left[  \omega_{1}\right]  & -ig_{b}^{a}\\
+ig_{a}^{b\ast} & h_{b}\left[  \omega_{2}\right]
\end{array}
\right]  \left[
\begin{array}
[c]{c}%
a^{\mathrm{in}}\left[  +\omega_{1}\right] \\
b^{\mathrm{in}}\left[  -\omega_{2}\right]
\end{array}
\right]  ,
\end{multline}
where
\begin{equation}
h_{a,b}\left[  \omega\right]  =-i\omega+i(\omega_{a,b}-i\Gamma_{a,b})
\end{equation}
and the signal and idler angular frequencies $\omega_{1}$ and $\omega_{2}$ are
both positive, satisfying the relationship
\begin{equation}
\omega_{1}+\omega_{2}=\omega_{c}.
\end{equation}
The scattering matrix of the device for small signals is defined by
\begin{equation}
\left[
\begin{array}
[c]{c}%
a^{\mathrm{out}}\left[  +\omega_{1}\right] \\
b^{\mathrm{out}}\left[  -\omega_{2}\right]
\end{array}
\right]  =\left[
\begin{array}
[c]{cc}%
r_{aa} & s_{ab}\\
s_{ba} & r_{bb}%
\end{array}
\right]  \left[
\begin{array}
[c]{c}%
a^{\mathrm{in}}\left[  +\omega_{1}\right] \\
b^{\mathrm{in}}\left[  -\omega_{2}\right]
\end{array}
\right]  . \label{RedS}%
\end{equation}
It can be computed from Eq. (\ref{in-out-amp5}) and one finds
\begin{align}
r_{aa}  &  =\frac{\chi_{a}^{-1\ast}\chi_{b}^{-1\ast}+\left\vert \rho
\right\vert ^{2}}{\chi_{a}^{-1}\chi_{b}^{-1\ast}-\left\vert \rho\right\vert
^{2}},\\
r_{bb}  &  =\frac{\chi_{a}^{-1}\chi_{b}^{-1}+\left\vert \rho\right\vert ^{2}%
}{\chi_{a}^{-1}\chi_{b}^{-1\ast}-\left\vert \rho\right\vert ^{2}},\\
s_{ab}  &  =\frac{-2i\rho}{\chi_{a}^{-1}\chi_{b}^{-1\ast}-\left\vert
\rho\right\vert ^{2}},\\
s_{ba}  &  =\frac{2i\rho^{\ast}}{\chi_{a}^{-1}\chi_{b}^{-1\ast}-\left\vert
\rho\right\vert ^{2}},
\end{align}
where the $\chi$'s are the bare response functions of modes \textit{a} and
\textit{b} (whose inverses depend linearly on the signal frequency)
\begin{align}
\chi_{a}^{-1}  &  =1-i\frac{\omega_{1}-\omega_{a}}{\Gamma_{a}},\\
\chi_{b}^{-1}  &  =1-i\frac{\omega_{2}-\omega_{b}}{\Gamma_{b}},
\end{align}
and $\rho$ is the dimensionless pump amplitude
\begin{equation}
\rho=\frac{g_{3}\sqrt{\bar{n}_{c}}e^{-i\phi}}{\sqrt{\Gamma_{a}\Gamma_{b}}}.
\label{rho}%
\end{equation}
Note that the matrix in Eq. (\ref{RedS}) has unity determinant and the
property
\begin{align}
\left\vert r_{aa}\right\vert ^{2}-\left\vert s_{ab}\right\vert ^{2}  &  =1,\\
\left\vert r_{bb}\right\vert ^{2}-\left\vert s_{ba}\right\vert ^{2}  &  =1.
\end{align}
For zero frequency detuning, i.e. $\chi_{a}^{-1}=\chi_{b}^{-1}=1$, the
scattering matrix displays a very simple form
\begin{equation}
\left[
\begin{array}
[c]{cc}%
\cosh\tau_{0} & -ie^{-i\phi}\sinh\tau_{0}\\
+ie^{+i\phi}\sinh\tau_{0} & \cosh\tau_{0}%
\end{array}
\right]  , \label{hyperbolic}%
\end{equation}
where $\tanh(\tau_{0}/2)=|\rho|$. The zero frequency detuning power gain
$G_{0}$ is given by
\begin{equation}
G_{0}=\left(  \cosh\tau_{0}\right)  ^{2}=\left(  \frac{1+\left\vert
\rho\right\vert ^{2}}{1-\left\vert \rho\right\vert ^{2}}\right)  ^{2}.
\label{gain-formula}%
\end{equation}
For non-zero detuning, the scattering matrix acquires extra phase factors but
the minimal scattering matrix for a quantum-limited phase-preserving amplifier
represented in Fig. \ref{minimal-amp} still describes the device.%
The gain $G_{0}$ diverges as $\left\vert \rho\right\vert \rightarrow1^{-}$,
i.e. when the photon number $\bar{n}_{c}$ in the pump resonator reaches the
critical number given by \newline%
\begin{equation}
\bar{n}_{c}^{po}=\frac{\Gamma_{a}\Gamma_{b}}{\left\vert g_{3}\right\vert ^{2}%
}, \label{onset-param-osc}%
\end{equation}
a result that is common to all forms of parametric amplification. Increasing
the pump power beyond the critical power yielding $\bar{n}_{c}^{po}$ leads to
the parametric oscillation regime. This phenomenon is beyond the scope of our
simple analysis and cannot be described by our starting equations, since
higher order non-linearities of the system need to be precisely modelled if
the saturation of the oscillation is to be accounted for.

Introducing the detuning
\begin{equation}
\Delta\omega=\omega_{1}-\omega_{a}=\omega_{b}-\omega_{2},
\end{equation}
we can give a useful expression for the gain as a function of frequency as
\begin{equation}
G\left(  \Delta\omega\right)  \underset{\left\vert \rho\right\vert
\rightarrow1^{-}}{=}\frac{G_{0}}{1+\left(  \frac{\Delta\omega}{\gamma
G_{0}^{-1/2}}\right)  ^{2}},
\end{equation}
which shows that in the limit of large gain, the response of the amplifier for
both the signal and idler port is Lorentzian with a bandwidth given by
\begin{equation}
B=2\gamma G_{0}^{-1/2}=\frac{2\gamma_{a}\gamma_{b}G_{0}^{-1/2}}{\gamma
_{a}+\gamma_{b}}. \label{B}%
\end{equation}
The product of the maximal amplitude gain times the bandwidth is thus constant
and is given by the harmonic average of the oscillator bandwidths. Another
interesting prediction of the scattering matrix is the two-mode squeezing
function of the device demonstrated in Ref. \onlinecite{NoiseCorrelation}.
%
%
\subsection{Conversion without photon gain (case 2)}
\begin{figure}
[h]
\begin{center}
\includegraphics[width=\columnwidth]{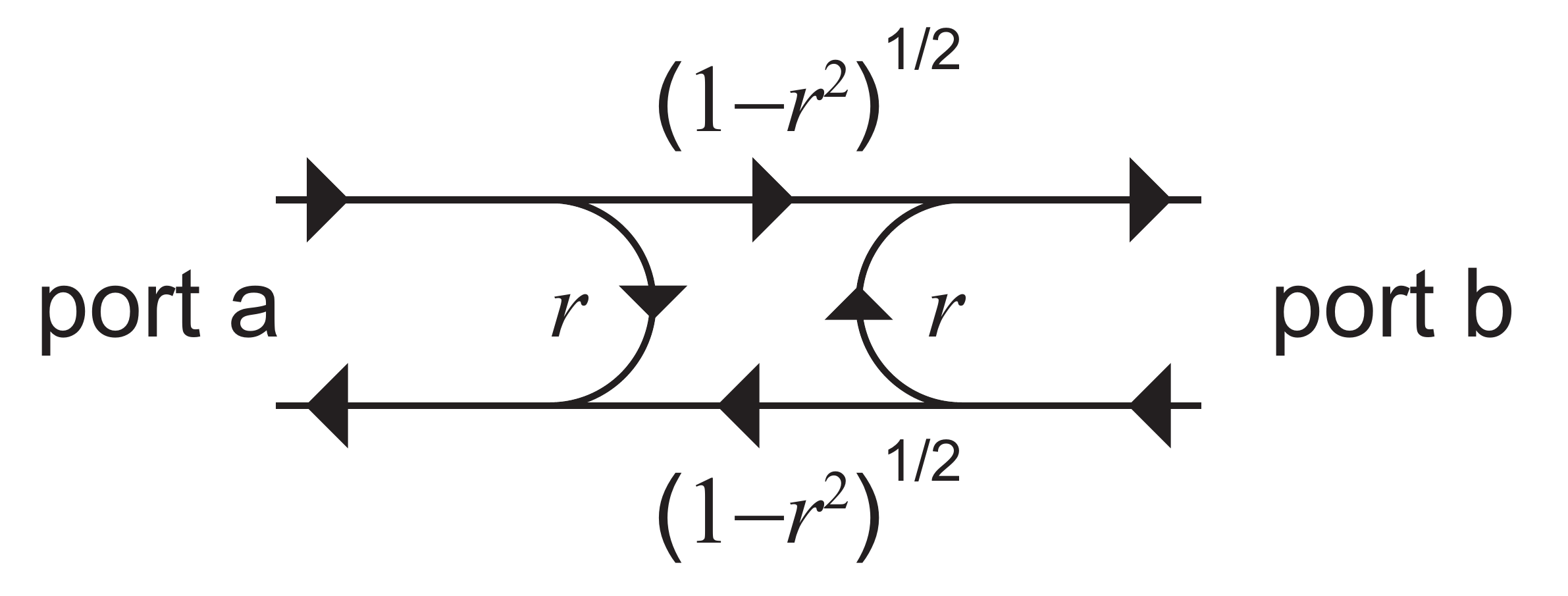}\\
\caption{Signal flow graph for a three-wave mixing device operating in
conversion without photon gain, realized in case 2 of Fig. \ref{frequencies}.
The incoming signal in port $a$ ($b$) is reflected with amplitude $r$ and
transmitted with up-conversion (down-conversion) to port $b$ (a) with
amplitude $(1-r^{2})^{1/2}$. }%
\label{convflow}%
\end{center}
\end{figure}

%
%
The case of conversion without photon gain can be treated along the same line
as in the previous subsection, where scattering takes place between \textit{c}
and \textit{a} or \textit{c} and \textit{b }modes. Without loss of generality
we assume that the pump is applied to the intermediate frequency resonance. In
this case the scattering matrix reads%

\begin{equation}
\left[
\begin{array}
[c]{c}%
a^{\mathrm{out}}\left[  +\omega_{1}\right] \\
c^{\mathrm{out}}\left[  +\omega_{3}\right]
\end{array}
\right]  =\left[
\begin{array}
[c]{cc}%
r_{aa} & t_{ac}\\
t_{ca} & r_{cc}%
\end{array}
\right]  \left[
\begin{array}
[c]{c}%
a^{\mathrm{in}}\left[  +\omega_{1}\right] \\
c^{\mathrm{in}}\left[  +\omega_{3}\right]
\end{array}
\right]  ,
\end{equation}
where%

\begin{align}
r_{aa}  &  =\frac{\chi_{a}^{-1\ast}\chi_{c}^{-1}-\left\vert \rho^{\prime
}\right\vert ^{2}}{\chi_{a}^{-1}\chi_{c}^{-1}+\left\vert \rho^{\prime
}\right\vert ^{2}},\nonumber\\
r_{cc}  &  =\frac{\chi_{a}^{-1}\chi_{c}^{-1\ast}-\left\vert \rho^{\prime
}\right\vert ^{2}}{\chi_{a}^{-1}\chi_{c}^{-1}+\left\vert \rho^{\prime
}\right\vert ^{2}},\nonumber\\
t_{ac}  &  =\frac{2i\rho^{\prime}}{\chi_{a}^{-1}\chi_{c}^{-1}+\left\vert
\rho^{\prime}\right\vert ^{2}},\nonumber\\
t_{ca}  &  =\frac{2i\rho^{\prime\ast}}{\chi_{a}^{-1}\chi_{c}^{-1}+\left\vert
\rho^{\prime}\right\vert ^{2}},\nonumber\\
&  \label{SparamsConvYSeries}%
\end{align}
and%

\begin{align}
\chi_{c}^{-1}  &  =1-i\frac{\omega_{3}-\omega_{c}}{\Gamma_{c}},\\
\rho^{\prime}  &  =\frac{g_{3}\sqrt{\bar{n}_{b}}e^{-i\phi}}{\sqrt{\Gamma
_{a}\Gamma_{c}}}.
\end{align}
The reduced pump strength $\rho^{\prime}$ plays the same role here as $\rho$
in the photon amplification case. Note that the scattering matrix is now
unitary (conservation of total number of photons) and satisfies the following relations:%

\begin{align}
\left\vert r_{aa}\right\vert ^{2}+\left\vert t_{ac}\right\vert ^{2}  &  =1,\\
\left\vert r_{cc}\right\vert ^{2}+\left\vert t_{ca}\right\vert ^{2}  &  =1.
\end{align}
For zero frequency detuning, i.e. $\chi_{a}^{-1}=\chi_{c}^{-1}=1$, the
scattering matrix can be written as%

\begin{equation}
\left[
\begin{array}
[c]{cc}%
\cos\tau_{0} & e^{-i\phi}\sin\tau_{0}\\
e^{i\phi}\sin\tau_{0} & \cos\tau_{0}%
\end{array}
\right]  ,
\end{equation}
which corresponds to replacing the parameter $\tau_{0}$ by $i\tau_{0}$ or
$\left\vert \rho\right\vert $ by $i\left\vert \rho\right\vert $ in the
scattering matrix (\ref{hyperbolic}). A scattering representation of the
two-port device in conversion mode is shown in Fig. \ref{convflow}. In this
mode the device operates as a beam splitter, the only difference being that
the photons in different arms have different frequencies \cite{BSconv}. Full conversion
$\left(  \sin\tau_{0}=1\right)  $ is obtained on resonance when the pump power
reaches the critical value. However, here, the critical value can be traversed
without violating the validity of the equations. Full photon conversion is
desirable in dynamical cooling: in that case, the higher frequency resonator
will be emptied of photons, and the lower frequency resonator can be cooled to
its ground state by pumping the intermediate frequency resonator (see lower
panel of Fig. \ref{frequencies}).%
%
%
\subsection{Added Noise}
%
%
The number of output photons generated per mode in the amplification (case 1)
is given by%

\begin{equation}
\mathcal{N}_{a,b}^{\mathrm{out}}=\left\vert r\right\vert ^{2}\mathcal{N}%
_{a,b}^{\mathrm{in}}+\left\vert s\right\vert ^{2}\mathcal{N}_{b,a}%
^{\mathrm{in}}, \label{Nab_out_ampl}%
\end{equation}
where $\mathcal{N}^{\mathrm{in}}$ is the input photon spectral density given
by Eq. (\ref{Na_in_sec}) and we assume that there is no cross-correlations
between the input fields $a^{\mathrm{in}}$ and $b^{\mathrm{in}}$.

Assuming that the three-wave mixing device is in thermal equilibrium at
temperature $T\ll\hbar\omega_{1,2}/k_{B}$ and that the dominant noise entering
the system at each port is zero-point fluctuations $\hbar\omega_{1,2}/2$
($\mathcal{N}^{\mathrm{in}}=1/2$), then in the limit of high gain $\left\vert
r\right\vert \gg1$, the number of noise equivalent photons effectively feeding
the system is%

\begin{equation}
\mathcal{N}_{eq}^{\mathrm{in}}=\mathcal{N}^{\mathrm{out}}/\left\vert
r\right\vert ^{2}\simeq1. \label{N_in_eff_refl}%
\end{equation}
This means that the number of noise equivalent photons added by the device to
the input is given by $\mathcal{N}^{\mathrm{add}}=\mathcal{N}_{eq}%
^{\mathrm{in}}-\mathcal{N}^{\mathrm{in}}=1/2$. Hence, when operated as a
non-degenerate amplifier with $G_{0}\gg1$, the device adds noise which is
equivalent to at least half a photon at the signal frequency to the input, in
agreement with Caves theorem \cite{Caves}.

In contrast, in the conversion mode of operation, assuming that there is no
correlation between the input fields, the number of generated output photons
per mode reads%

\begin{equation}
\mathcal{N}_{a,b}^{\mathrm{out}}=\left\vert r\right\vert ^{2}\mathcal{N}%
_{a,b}^{\mathrm{in}}+\left\vert t\right\vert ^{2}\mathcal{N}_{b,a}%
^{\mathrm{in}}.
\end{equation}
Therefore, in pure conversion where $\left\vert r\right\vert =0$ and
$\left\vert t\right\vert =1$, when referring the noise back to the input, one
gets noise equivalent photons%

\begin{equation}
\mathcal{N}_{eq}^{\mathrm{in}}=\mathcal{N}^{\mathrm{out}}/\left\vert
t\right\vert ^{2}=1/2. \label{N_in_eff_pure}%
\end{equation}
This means that, as a converter, the device is not required to add noise to
the input since $\mathcal{N}_{eq}^{\mathrm{in}}=\mathcal{N}^{\mathrm{in}}$.
%
%
\section{Three-wave mixing using JRM}
\begin{figure}
[h]
\begin{center}
\includegraphics[width=\columnwidth]{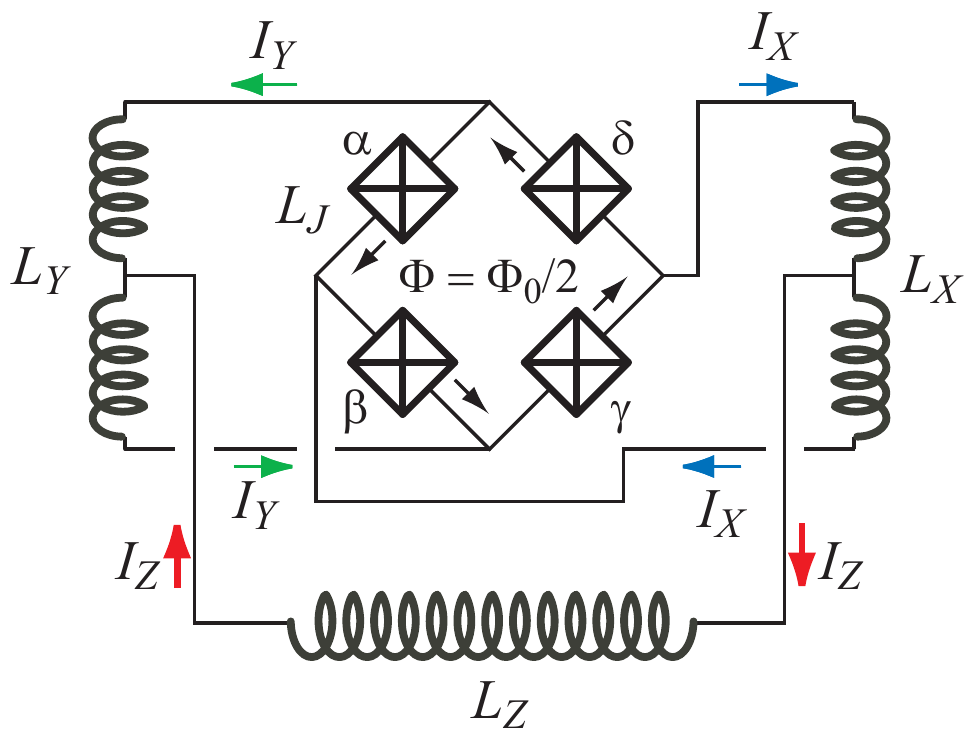}\\
\caption{Three-wave mixing element (see ellipse marked $K$ in Fig.
\ref{three_osc_fig}) consisting of a loop of four nominally identical
Josephson junctions threaded by a flux in the vicinity of half a flux quantum.
Mutual inductances, not shown here, couple this circuit to inductances $L_{a}%
$, $L_{b}$ and $L_{c}$ of Fig. \ref{three_osc_fig} via the inductances $L_{X}%
$, $L_{Y}$ and $L_{Z}$ respectively, which are much larger than the junction
inductance $L_{J}$. The three currents $I_{X}$, $I_{Y}$ and $I_{Z}$ correspond
to the three orthogonal modes of the structure.}%
\label{Josephson_ring_modulator}%
\end{center}
\end{figure}

%
%
The Josephson ring modulator is a device consisting of four Josephson
junctions, each with critical current $I_{0}=\frac{\hbar}{2eL_{J}}$ forming a
ring threaded by a flux $\Phi=\Phi_{0}/2$ where $\Phi_{0}$ is the flux quantum
(see Fig. \ref{Josephson_ring_modulator}). The device has the symmetry of a
Wheatstone bridge.%
There are thus three orthogonal electrical modes coupled to the junctions,
corresponding to the currents $I_{X}$, $I_{Y}$ and $I_{Z}$ flowing in three
external inductances $L_{X}$, $L_{Y}$ and $L_{Z}$ that are much larger than
the junction inductance $L_{J}=\varphi_{0}^{2}E_{J}^{-1}$, where $\varphi
_{0}=\hbar/2e$ is the reduced flux quantum. Each junction $j\in\left\{
\alpha,\beta,\gamma,\delta\right\}  $ is traversed by a current $I_{j}$ and at
the working point (i.e. $\Phi=\Phi_{0}/2$) its energy is, keeping terms up to
order four in $I_{j}$, given by
\begin{equation}
E_{j}=\frac{1}{2}L_{J}^{eff}I_{j}^{2}-\frac{1}{24}\frac{L_{J}^{eff}}%
{I_{0}^{\prime2}}I_{j}^{4},
\end{equation}
where $L_{J}^{eff}=\sqrt{2}L_{J}$ and $I_{0}^{\prime}=I_{0}/\sqrt{2}$. The
currents in the junctions are expressed by
\begin{align}
I_{\alpha}  &  =\frac{-I_{X}-I_{Y}}{2}+\frac{I_{Z}}{4}+I_{\Phi},\\
I_{\beta}  &  =\frac{+I_{X}-I_{Y}}{2}-\frac{I_{Z}}{4}+I_{\Phi},\\
I_{\gamma}  &  =\frac{+I_{X}+I_{Y}}{2}+\frac{I_{Z}}{4}+I_{\Phi},\\
I_{\delta}  &  =\frac{-I_{X}+I_{Y}}{2}-\frac{I_{Z}}{4}+I_{\Phi},
\end{align}
where $I_{\Phi}$ is the supercurrent induced in the ring by the externally
applied flux $\Phi$. The total energy of the ring is, keeping terms up to
third order in the currents \cite{HuardProc},
\begin{equation}
E_{ring}=\frac{1}{2}L_{J}^{eff}\left(  I_{X}^{2}+I_{Y}^{2}+\frac{1}{4}%
I_{Z}^{2}\right)  -\frac{1}{4}\frac{L_{J}^{eff}I_{\Phi}}{I_{0}^{\prime2}}%
I_{X}I_{Y}I_{Z}.
\end{equation}
We can express the currents as
\begin{equation}
I_{X,Y,Z}=\frac{\Phi_{a,b,c}}{L_{a,b,c}}\frac{M_{a,b,c}}{L_{X,Y,Z}}=\frac
{\Phi_{a,b,c}}{L_{a,b,c}^{eff}},
\end{equation}
where $M_{a,b,c}$ are the mutual inductances between $L_{X,Y,Z}$ and the
oscillator inductances $L_{a,b,c}$. The non-linear coefficient in the energy
is, therefore,
\begin{equation}
K=\frac{\left(  L_{J}^{eff}\right)  ^{2}}{4\varphi_{0}}\frac{1}{L_{a}%
^{eff}L_{b}^{eff}L_{c}^{eff}},
\end{equation}
and we finally arrive at the result
\begin{equation}
g_{3}^{2}=\frac{p_{a}p_{b}p_{c}\omega_{a}\omega_{b}\omega_{c}}{\omega
_{J}^{eff}}. \label{eqng3}%
\end{equation}
Here the participation ratios are defined as
\begin{equation}
p_{a,b,c}=\frac{L_{J}^{eff}}{L_{a,b,c}^{eff}},
\end{equation}
and, at $\Phi=\Phi_{0}/2$,
\begin{equation}
\omega_{J}^{eff}=\frac{128}{\sqrt{2}}\frac{E_{J}}{\hbar}.
\end{equation}
The participation ratios are linked to the maximal number of photons in each
resonator, defined as those corresponding to an oscillation amplitude reaching
a current of $I_{0}$ in each junction of the ring modulator,
\begin{equation}
p_{a,b,c}\bar{n}_{a,b,c}^{\max}=\frac{E_{J}^{a,b,c}}{\hbar\omega_{a,b,c}},
\label{p-min}%
\end{equation}
where the $E_{J}^{a,b,c}$ are of order $E_{J}$ with factors accounting for the
different participation of modes $X,$ $Y$ and $Z$ in the current of each
junction. Equations (\ref{eqng3}) and (\ref{p-min}) are valid for all types of
coupling between the Josephson ring modulator and signal/pump oscillators,
which can be realized in practice by inductance sharing rather than by the
mutual inductances discussed here.

Equation (\ref{p-min}) can also be rewritten in terms of the maximum
circulating power in cavities \textit{a} and \textit{b} as%

\begin{equation}
P_{\mathrm{cav}}^{\mathrm{\max}}=\frac{\gamma_{a,b}}{p_{a,b}}\frac{E_{J}%
}{\sqrt{2}} \label{Pmaxcav}%
\end{equation}
where we substituted $E_{J}/\sqrt{2}$ as an upper bound for $E_{J}^{a,b}$. The
maximum number of photons in equation (\ref{p-min}) determine the maximum
signal input power handled by the device
\begin{equation}
P_{a,b}^{\max}=\frac{1}{G}\gamma_{a,b}\hbar\omega_{a,b}\bar{n}_{a,b}^{\max}.
\label{P-Max}%
\end{equation}
We can now combine the notion of maximum power in resonator $c$ compatible
with weak non-linearity with that of a critical power for the onset of
parametric oscillation given by Eq. (\ref{onset-param-osc}):
\begin{equation}
\bar{n}_{c}^{\max}=\frac{E_{J}^{c}}{p_{c}\hbar\omega_{c}}>\bar{n}%
_{c}^{\mathrm{po}}=\frac{\Gamma_{1}\Gamma_{2}}{g_{3}^{2}},
\end{equation}
arriving at the important relation
\begin{equation}
p_{a}p_{b}Q_{a}Q_{b}>\Xi, \label{PQ-product}%
\end{equation}
where $\Xi$ is a number of order unity depending on the exact implementation
of the coupling between the ring modulator and the oscillators. The quality
factors of the resonators obey the well-known relation
\begin{equation}
Q_{a,b}=\frac{\omega_{a,b}}{\gamma_{a,b}}.
\end{equation}

Another maximum limit on the gain of the amplifier is set by the saturation of
the device due to amplified zero-point fluctuations present at the input given by%

\begin{equation}
G_{\mathrm{ZPF}}^{\max}=\frac{E_{J}}{\sqrt{2}p_{a,b}}\frac{2}{\hbar
\omega_{a,b}}. \label{Gmaxzpf}%
\end{equation}
Eqs. (\ref{p-min}), (\ref{P-Max}) and (\ref{PQ-product}) show that it is not
possible to maximize simultaneously gain, bandwidth and dynamic
range.\begin{table}[ptb]%
\begin{tabular}
[c]{|c|c|}\hline
\; \textrm{Parameter} & \; \textrm{Range}\\\hline\hline
$\omega_{a,b}/2\pi$ & 1 - 16\;\textrm{GHz}\\\hline
$Q_{a,b}$ & 50 - 500\\\hline
$Z_{a,b} $ & 10 - 150 $\Omega$\\\hline
$\gamma_{c} $ & 0.5 - 10\;\textrm{GHz}\\\hline
$I_{0} $ & 0.5 - 10 $\mu$\;\textrm{A}\\\hline
$E_{J} $ & 10 - 230\;\textrm{K}\\\hline
$p_{a,b,c} $ & 0.01 - 0.5\\\hline
$g_{3}/2\pi$ & 0.1 - 15\;\textrm{MHz}\\\hline
$\bar{n}^{\;\mathrm{max}}_{a,b,c} $ & $20 - 10^{4}$\\\hline
\end{tabular}
\label{Table2}\caption{Typical values for Josephson three-wave mixing devices.}%
\end{table}

In table II we enlist general bounds on the characteristic parameters of the
three-wave mixing device, which are feasible with superconducting microwave
circuits and standard Al-AlOx-Al junction fabrication technology. A few
comments regarding the values listed in the table are in order. The frequency
ranges of resonators \textit{a} and \textit{b} is mainly set by the center
frequency of the system whose signal one needs to amplify or process. It is
also important that these frequencies are very small compared to the plasma
frequency of the Josephson junction. The total quality factor range listed in
the table $\left(50-500\right)$ is suitable for practical devices. Quality
factors in excess of $500$ can be easily achieved with superconducting
resonators but, as seen from Eq. (\ref{B}), higher the quality factor, smaller
the dynamical bandwidth of the device. Quality factors lower than $50$ on the
other hand are not recommended either for a variety of reasons. For example,
in the limit of very low $Q$ the pump softens (becomes less stiff), and the
dynamic range decreases as more quantum noise will be admitted by the device
bandwidth and amplified ``unintentionally" by the junctions. The characteristic
impedance of the resonators $Z_{a,b}$ is set by microwave engineering
considerations as discussed in Sec. IV but, in general, this value varies
around 50 $\Omega$. The rate $\gamma_{c}$ at which pump photons leave the circuit varies from
one circuit design to the other as discussed in Sec. IV and is limited by
$\omega_{c}$. This parameter also affects the maximum input power performance
of the device as explained in Sec. III. As to the values of $I_{0}$, on the
one hand it is beneficial to work with large Josephson junctions in order to
increase the processing capability of the device; on the other hand a critical
current larger than 10 $\mu$A adds complexity to the microwave design of the resonators and makes the
fabrication process of the Josephson junction more involved. This might even
require switching to a different fabrication process such as Nb-AlOx-Nb
trilayer junctions \cite{Trilayer} or nanobridges \cite{Nanobridges}. The other parameters listed in the table,
namely $p_{a,b,c}$, $E_{J}$, $g_{3}$, $\bar{n}_{a,b,c}^{\max}$, their values
depend, to a large extent, on the device parameters already discussed.
%
%
\section{Limitation of dynamic range due to pump depletion}
%
%
In the last two sections, we were using results obtained by solving only the
first two of the equations of motion Eqs. (\ref{threeAmpEqs}) under the
restriction of the stiff pump approximation. In this section, we extend our
analysis and include the third equation describing the dynamics of the pump to
calculate the pump depletion and its effect on the dynamic range of the
device. For this purpose, we consider the average value of the third equation
of motion for field $c$
\begin{equation}
\frac{\mathrm{d}}{\mathrm{d}t}\left\langle c\right\rangle =-i\omega
_{c}\left\langle c\right\rangle -ig_{3}\left\langle ab\right\rangle
-\frac{\gamma_{c}}{2}\left\langle c\right\rangle +\sqrt{\gamma_{c}%
}\left\langle \tilde{c}^{\mathrm{in}}\left(  t\right)  \right\rangle .
\end{equation}
In steady state and using RWA we obtain
\begin{equation}
ig_{3}\left\langle ab\right\rangle +\frac{\gamma_{c}}{2}\left\langle c\left(
t\right)  \right\rangle =\sqrt{\gamma_{c}}\left\langle \tilde{c}^{\mathrm{in}%
}\left(  t\right)  \right\rangle . \label{cin_c_relation}%
\end{equation}
In the limit of vanishing input, the cross-correlation term $\left\langle
ab\right\rangle $ is negligible and, therefore,
\begin{equation}
\left\langle c\left(  t\right)  \right\rangle =\frac{2}{\sqrt{\gamma_{c}}%
}\left\langle \tilde{c}^{\mathrm{in}}\left(  t\right)  \right\rangle .
\end{equation}
The average number of photons in the $c$ resonator in this case is, thus,
\begin{equation}
\underset{\left\langle ab\right\rangle \rightarrow0}{\lim}\bar{n}_{c}=\frac
{4}{\gamma_{c}}\left\vert \left\langle \tilde{c}^{\mathrm{in}}\left(
t\right)  \right\rangle \right\vert ^{2}. \label{c_sq_c_in_sq}%
\end{equation}
We now establish a self-consistent equation for $\bar{n}_{c}$, taking into
account input signals of finite amplitude. We first evaluate the value of
$\left\langle a\left(  t\right)  b\left(  t\right)  \right\rangle $ in the
frame rotating with the pump phase,
\begin{align}
&  \left\langle a\left(  t\right)  b\left(  t\right)  \right\rangle
\nonumber\\
&  =\frac{1}{2\pi}\int_{-\infty}^{+\infty}\int_{-\infty}^{+\infty}\left\langle
a\left[  \omega\right]  b\left[  \omega^{\prime}\right]  \right\rangle
e^{-i\left(  \omega+\omega^{\prime}\right)  t}\mathrm{d}\omega\mathrm{d}%
\omega^{\prime}.
\end{align}
Using the field relations (see Appendix)
\begin{align}
\sqrt{\gamma_{a}}a\left[  \omega\right]   &  =\tilde{a}^{\mathrm{in}}\left[
\omega\right]  +\tilde{a}^{\mathrm{out}}\left[  \omega\right]  ,\\
\sqrt{\gamma_{b}}b\left[  \omega\right]   &  =\tilde{b}^{\mathrm{in}}\left[
\omega\right]  +\tilde{b}^{\mathrm{out}}\left[  \omega\right]
\end{align}
and the input-output relations given by Eq. (\ref{RedS}), we obtain
(transforming back into the time domain)
\begin{equation}
-ig_{3}\left\langle a\left(  t\right)  b\left(  t\right)  \right\rangle
=-\frac{\gamma_{eff}\left(  G\right)  }{2}\left\langle c\left(  t\right)
\right\rangle ,
\end{equation}
where, in the limit of large gains $G\gg1$,
\begin{equation}
\gamma_{eff}\left(  G\right)  =\frac{1}{2\pi}\frac{\gamma_{c}}{4\overline
{n}_{c}^{\mathrm{in}}}\int_{0}^{+\infty}\mathrm{d}\omega\left(  \mathcal{N}%
_{a}^{\mathrm{in}}\left[  \omega\right]  +\mathcal{N}_{b}^{\mathrm{in}}\left[
\omega\right]  \right)  G\left(  \Delta\omega\right)  \label{gamma_eff}%
\end{equation}
denotes an effective decay rate of pump photons due to generation of entangled
signal and idler photons. This last relation expresses, in another form, the
Manley-Rowe relations \cite{ManleyProc} that establish the equality between
the number of created signal photons by the amplifier to the number of
destroyed pump photons. It shows that even in the absence of any deterministic
signal applied to the oscillator $a$ or $b$, pump photons are used to amplify
zero-point fluctuations. Therefore, the pump tone always encounters a
dissipative load even when no signals are injected into the device.

For a continuous wave (CW) input power sent at the center frequency of the $a$
or $b$ oscillator, or both, we have
\begin{equation}
\gamma_{eff}\left(  G_{0},P^{\mathrm{in}}\right)  =\frac{\gamma_{c}%
}{4\overline{n}_{c}^{\mathrm{in}}}G_{0}P^{\mathrm{in}}, \label{gamma_eff1}%
\end{equation}
where $P^{\mathrm{in}}=P_{a}^{\mathrm{in}}+P_{b}^{\mathrm{in}}$ is given in
units of photon number per unit time and, in steady state,
\begin{equation}
\overline{n}_{c}\left(  G_{0},P^{\mathrm{in}}\right)  =\frac{4\gamma_{c}%
}{\left(  \gamma_{c}+\gamma_{eff}\left(  G_{0},P^{\mathrm{in}}\right)
\right)  ^{2}}\overline{n}_{c}^{\mathrm{in}}. \label{nc_ncin}%
\end{equation}
As a finite input power is applied to the signal oscillators, oscillator $c$
depopulates and, keeping the pump power constant, we get
\begin{align}
\frac{\overline{n}_{c}\left(  G_{0},P^{\mathrm{in}}\right)  }{\overline{n}%
_{c}\left(  G_{0},P^{\mathrm{in}}=0\right)  }  &  =\frac{1}{\left(
1+\frac{G_{0}P^{\mathrm{in}}}{4\overline{n}_{c}^{\mathrm{in}}}\right)  ^{2}}\\
&  \simeq1-\frac{G_{0}P^{\mathrm{in}}}{2\overline{n}_{c}^{\mathrm{in}}}.
\end{align}
On the other hand, from Eqs. (\ref{rho}) and (\ref{gain-formula}), the left
hand side is given by
\begin{equation}
\frac{\overline{n}_{c}\left(  G_{0},P^{\mathrm{in}}\right)  }{\overline{n}%
_{c}\left(  G_{0},P^{\mathrm{in}}=0\right)  }=\frac{\frac{\sqrt{G}-1}{\sqrt
{G}+1}}{\frac{\sqrt{G_{0}}-1}{\sqrt{G_{0}}+1}},
\end{equation}
where $G$ denotes the gain in the presence of $P_{\mathrm{in}}$. In the large
gain limit, if we fix the maximum decrease of gain due to pump depletion to
be
\begin{equation}
\frac{G}{G_{0}}>1-\varepsilon
\end{equation}
with $\varepsilon\ll1$, then we obtain
\begin{equation}
\frac{P^{\mathrm{in}}}{2\overline{n}_{c}^{\mathrm{in}}}<\varepsilon
G_{0}^{-3/2}, \label{pumpDepletion1}%
\end{equation}
which can also be rewritten as%

\begin{equation}
\frac{2\overline{n}_{c}^{\mathrm{in}}}{G_{0}P^{\mathrm{in}}}>\varepsilon
^{-1}\sqrt{G_{0}}.
\end{equation}
This relation shows that the ratio of the power of the pump tone to that of
the signal at the output of the amplifier must always be much larger than the
amplitude gain, in order for the linearity of the amplifier not to be
compromised by pump depletion effects.

\begin{figure}
[h]
\begin{center}
\includegraphics[width=\columnwidth]{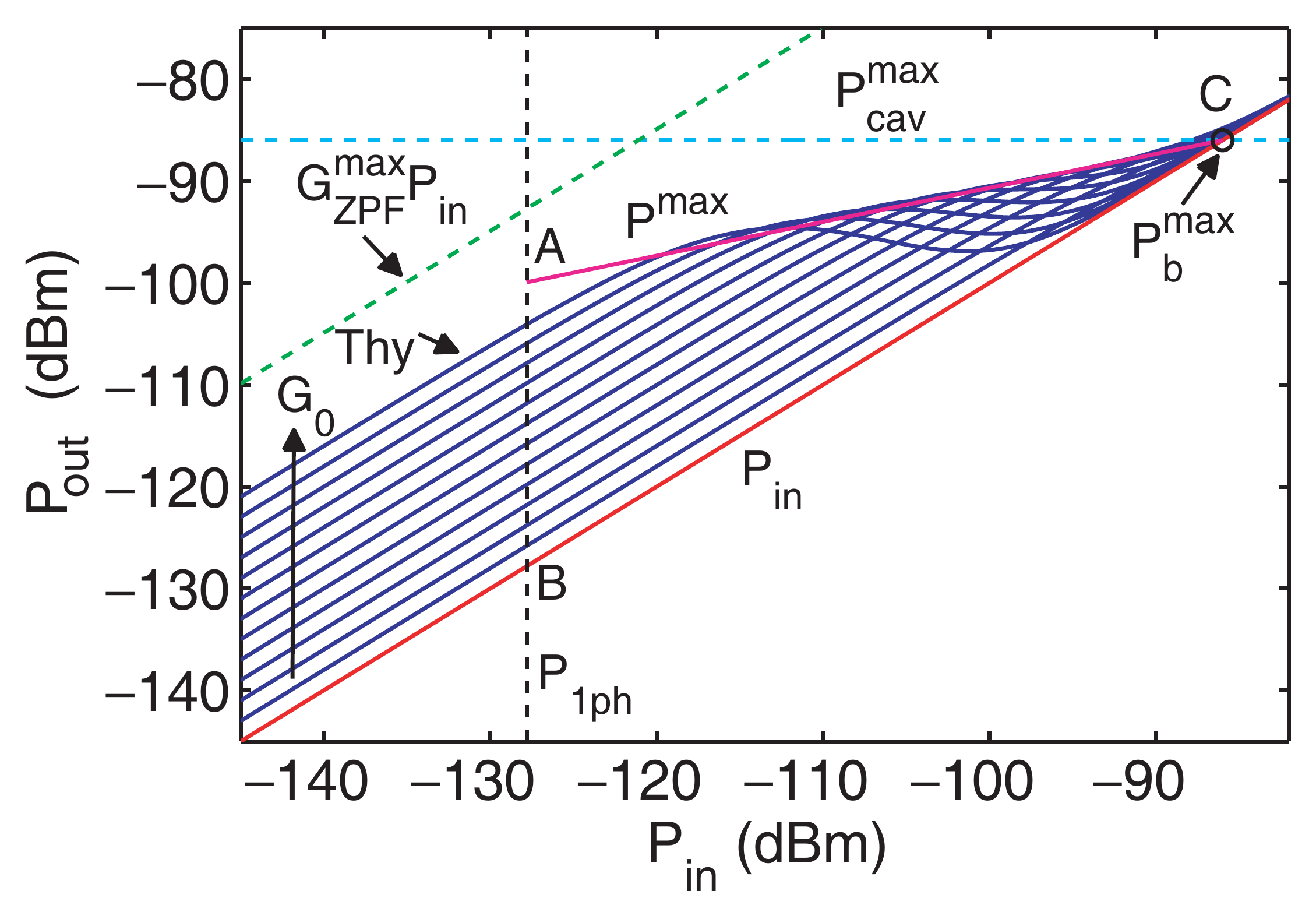}\\
\caption{(Color online). A calculated response of the signal output power
$P_{\mathrm{out}}$ versus the signal input power $P_{\mathrm{in}}$ of a
typical three-wave mixing device which exhibits a pump depletion effect. The
different blue curves correspond to different $G_{0}$ setpoints. The
definition of the other lines in the figure is given in the text. The
parameters used in the calculation are: $\omega_{a}/2\pi=7\operatorname{GHz}$,
$\omega_{b}/2\pi=8\operatorname{GHz}$, $\omega_{c}/2\pi=15\operatorname{GHz}$,
$\gamma_{a}/2\pi=\gamma_{b}/2\pi=50\operatorname{MHz}$, $\gamma_{c}%
/2\pi=0.6\operatorname{GHz}$, $Q_{a}=140$, $Q_{b}=160,$ $p_{a}=p_{b}=0.03$,
$p_{c}=0.02,$ $I_{0}=1\operatorname{\mu A}$, $E_{J}^{a,b}=E_{J}/\sqrt
{2}=16.3\operatorname{K}$, $P_{\mathrm{1ph}}=-128$ dBm, $G_{\mathrm{ZPF}%
}^{\max}=35$ dB, $P_{\mathrm{cav}}^{\mathrm{\max}}=P_{b}^{\mathrm{\max}}=-86$
dBm and $g_{3}/2\pi=0.7\operatorname{MHz}$.}%
\label{drthyfig}%
\end{center}
\end{figure}

In Fig. \ref{drthyfig} we plot a calculated response of the signal output
power $P_{\mathrm{out}}$ versus the signal input power $P_{\mathrm{in}}$ for a
typical three-wave mixing device. The device parameters employed in the
calculation and listed in the figure caption are practical values yielding a
maximum input power, which is limited by the effect of pump depletion. The
different blue curves are obtained by solving Eq. (\ref{nc_ncin}) for $G$ and
using the input-output relation $P_{\mathrm{out}}=GP_{\mathrm{in}}$, where
$P_{\mathrm{in}}$ expressed in units of power is taken as the independent
variable and $G_{0}$ is treated as a parameter. Note that in solving Eq.
(\ref{nc_ncin}), equations (\ref{gamma_eff1}), (\ref{c_sq_c_in_sq}),
(\ref{rho}) and (\ref{gain-formula}) are used. When drawn on logarithmic
scale, the device gain translates into a vertical offset (arrow indicating
$G_{0}$) off the $P_{\mathrm{out}}=P_{\mathrm{in}}$ line, indicated in red.
The dashed black vertical line corresponds to a signal input power of 1 photon
at the signal frequency per inverse dynamical bandwidth of the device at
$G_{0}=20$ dB. The dashed green line corresponds to the maximum gain set by
the amplified zero-point fluctuations given by Eq. (\ref{Gmaxzpf}), while the
cyan line corresponds to the maximum circulating power in the cavity given by
Eq. (\ref{Pmaxcav}).%
Furthermore, the maximum bound $P^{\max}$ indicated by the solid magenta line
corresponds to $P_{\mathrm{out}}^{\max}=G_{0}P_{\mathrm{in}}^{\max}$, where
$P_{\mathrm{in}}^{\max}=P_{b}^{\max}/G_{0}^{3/2}$ and $P_{b}^{\max
}=P_{\mathrm{cav}}^{\mathrm{\max}}$. As can be seen in the figure the
predicted power scaling due to pump depletion effect, expressed in relation
(\ref{pumpDepletion1}), follows the calculated response quite well. Finally,
it is straightforward to see that the usable region in the parameter space of
the device with respect to gain, bandwidth and maximum input power lies within
the boundaries of the fictitious triangle ABC indicated in the figure which is
formed by the intersection of the magenta, black and red lines.
%
%
\section{The Josephson parametric converter}
%
%
We discuss here three different realizations of the Josephson parametric
converter (JPC), which constitutes a fully non-degenerate three-wave mixing
device capable of amplification and conversion as discussed in the previous
sections. The three schemes differ in the resonator circuit design and the
coupling between the feedline and the resonator.
%
%
\subsection{Microstrip Resonator JPC (MRJ)}%
%
\begin{figure}
[h]
\begin{center}
\includegraphics[width=\columnwidth]{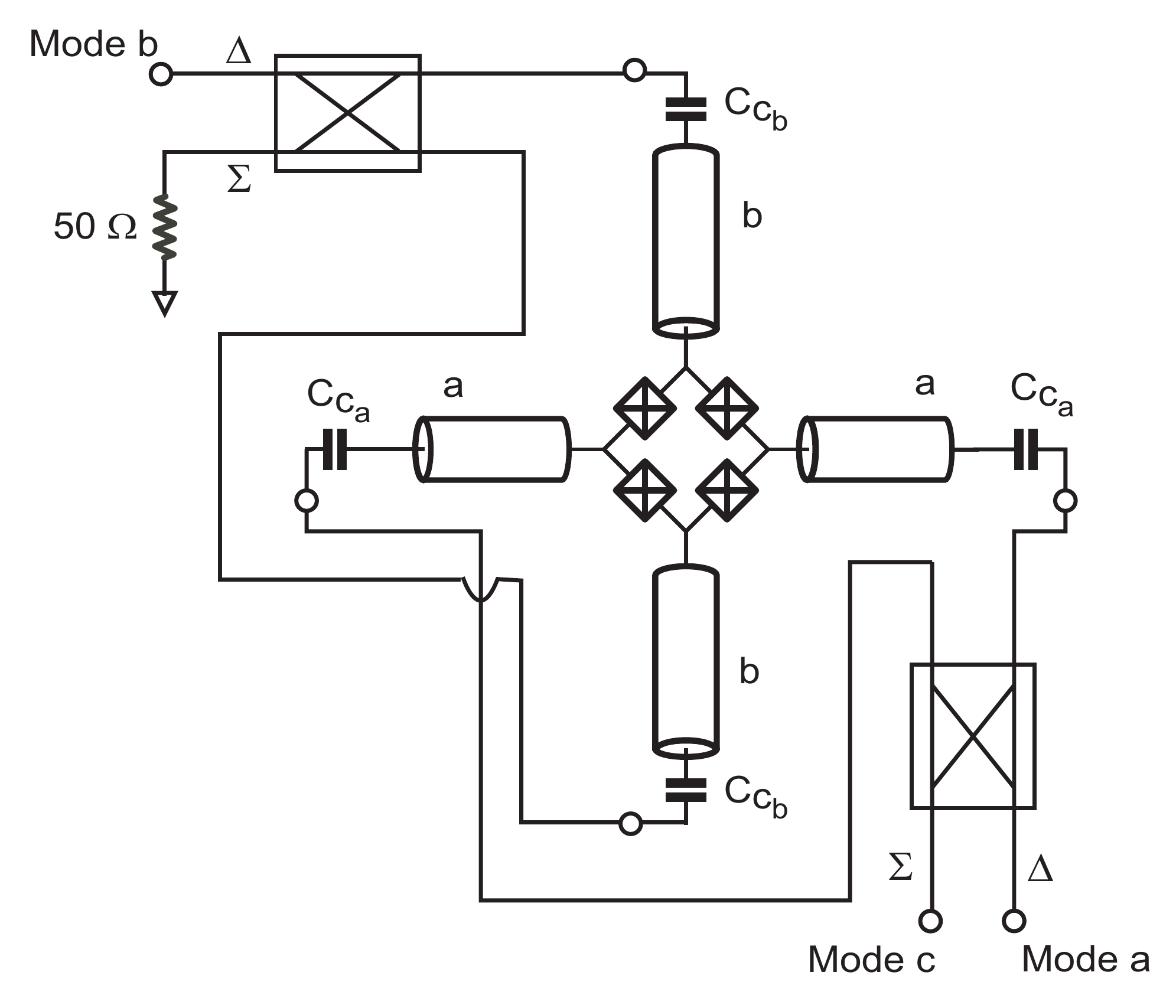}\\
\caption{Circuit model of the Microstrip Resonator JPC (MRJ).}%
\label{MJcirc}%
\end{center}
\end{figure}
\begin{figure}
[h]
\begin{center}
\includegraphics[width=\columnwidth]{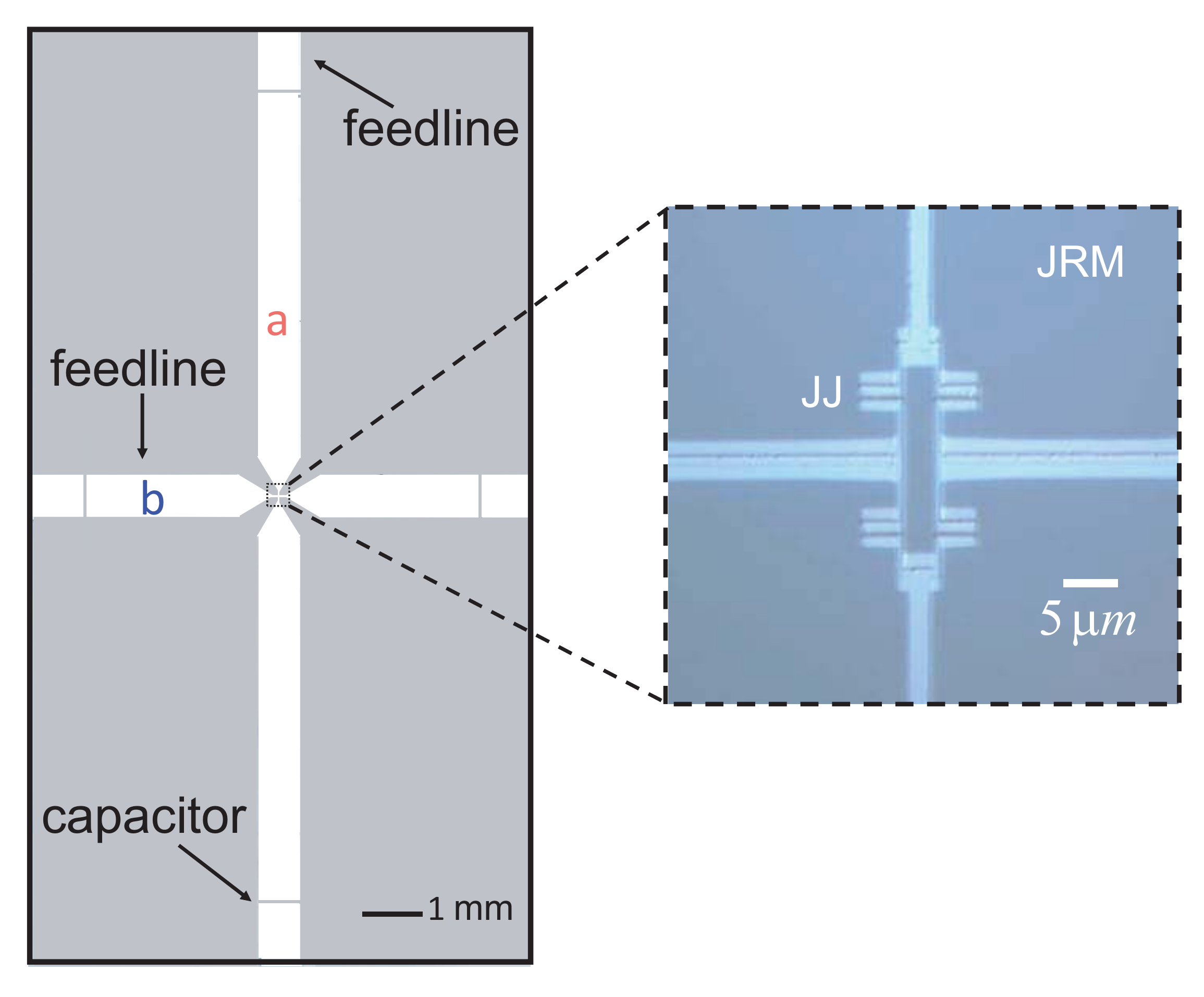}\\
\caption{(Color online). Optical microscope image of a microstrip resonator
JPC (MRJ). The resonators denoted \textit{a} and \textit{b }are half-wave
microstrip resonators which intersect at a JRM. A zoomed-in view of the
JRM, which consists of four Josephson junctions arranged in Wheatstone bridge
configuration, is shown on the right. The MRJ is coupled to
50 $\Omega$ feedlines via gap capacitors.}%
\label{MJphoto}%
\end{center}
\end{figure}

The MRJ comprises two superconducting microstrip resonators which intersect at
a JRM at the center as shown in the circuit model of the device in Fig.
\ref{MJcirc}. The resonance frequencies of the MRJ are determined by the
lengths of the microstrips $l_{a}\simeq\lambda_{a}/2$ and $l_{b}\simeq
\lambda_{b}/2$ and the Josephson inductance of the JRM, where $\lambda_{a}$,
$\lambda_{b}$ are the wavelengths of the fundamental resonances at $\omega
_{a}$ and $\omega_{b}$. It is worth mentioning that in addition to the
differential modes \textit{a} and \textit{b}, this configuration of two
coupled resonators also supports a common (even) mode. The angular frequency
$\omega_{e}$ at which this even mode resonates lies between $\left(
\omega_{b}+\omega_{a}\right)  /2$ and $\omega_{b}$ (where $\omega_{b}%
>\omega_{a}$). The characteristic impedance of the resonators in the MRJ model
is designed to be 50 $\Omega$ to ensure optimal coupling to the feedlines. Figure \ref{MJphoto} exhibits
an optical image of a typical MRJ device. The resonators are usually made of
Al or Nb over sapphire or high-resistivity silicon and are coupled to the
(transmission-line) feedlines using gap capacitors. The main role of these
coupling capacitors is to set the external quality factor of the resonators.
For a large bandwidth device operating in the $6-10$ GHz band, the external $Q$ of the resonators is typically in the range $60-100$.
In all JPC designs discussed here the total $Q$ essentially coincides with the
external $Q$, since the internal losses of the resonators are less than
$10^{-4}$. Signals at $\omega_{1}$ and $\omega_{2}$, which lie within the
bandwidths of resonators \textit{a} and \textit{b}, are fed into the JPC
through the delta port of a $180$ degree hybrid, whereas the pump drive
applied at $\omega_{3}=\omega_{1}+\omega_{2}$, for amplification, is a
non-resonant tone and is injected into the device through the sigma port of
the hybrid (Fig. \ref{MJcirc}). The main advantage of the MRJ is that it is
easy to design and fabricate. On the other hand, the main disadvantages are:
(1) the area of the device can be relatively large depending on the
frequencies of interest, (2) the characteristic impedance of the device is
limited to around 50 $\Omega$, (3) the pump can be less stiff than the designs discussed below. The latter
is due to the fact that the transmission-line resonators support higher
resonance modes such as $2\omega_{a}$ and $2\omega_{b}$ with finite $Q$, which
can be relatively close to the pump angular frequency $\omega_{3}$.%
%
%
\subsection{Compact Resonator JPC (CRJ)}%
%
%
\begin{figure}
[h]
\begin{center}
\includegraphics[width=\columnwidth]{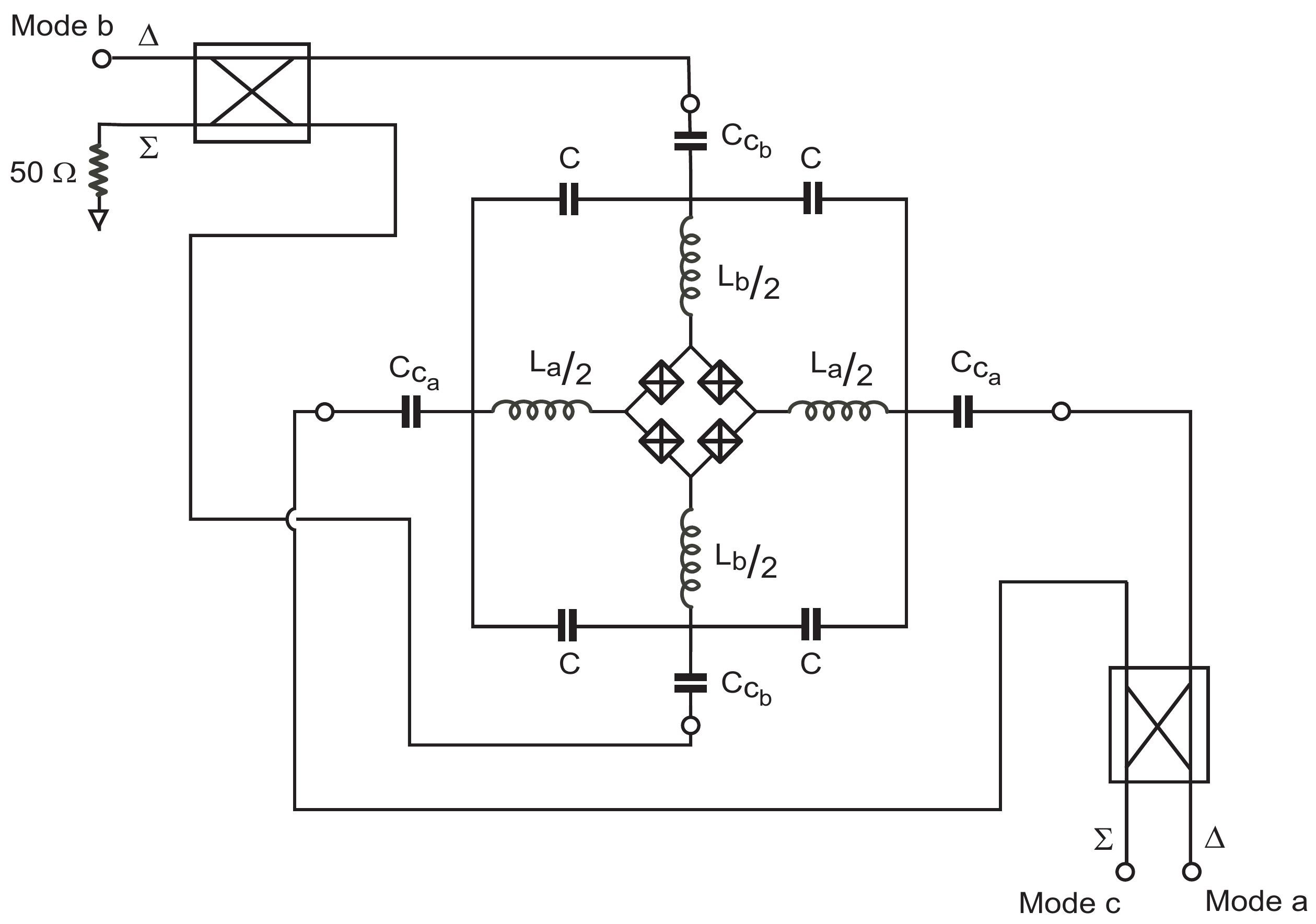}\\
\caption{Circuit model of the Compact Resonator JPC (CRJ).}%
\label{CJcirc}%
\end{center}
\end{figure}
\begin{figure}[h]
\begin{center}
\includegraphics[width=\columnwidth]{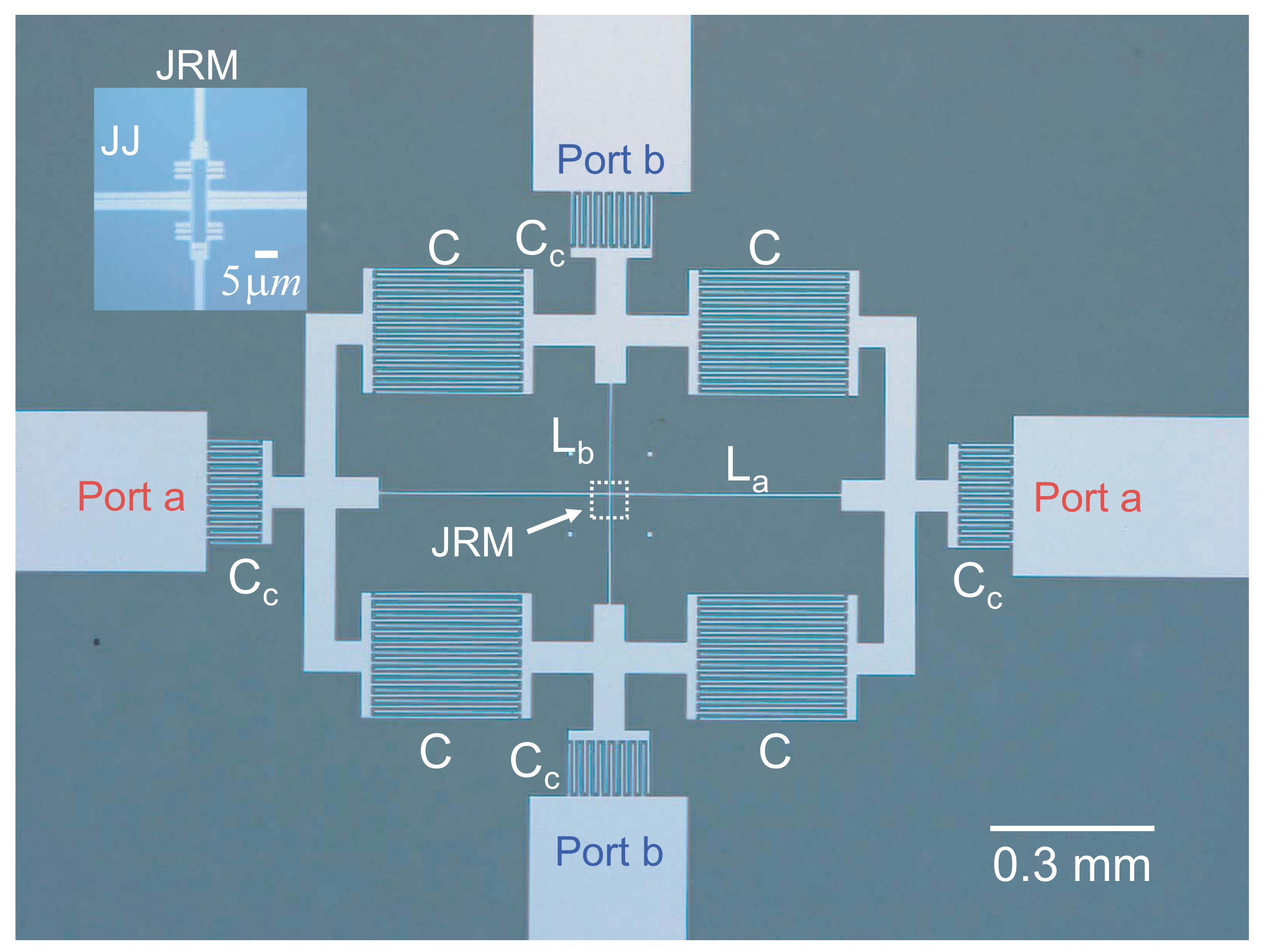}\\
\caption{(Color online). Optical microscope image of a compact resonator JPC
(CRJ). The device consists of four equal interdigitated capacitors denoted $C$
and two inductive elements denoted $L_{a}$ and $L_{b}$ which are realized
using narrow superconducting lines of different lengths. The JRM of the device
resides at the intersection of the two lines. An optical image of the JRM is
shown in the inset. The CRJ is coupled to $50\operatorname{\Omega }$
microstrip feedlines via interdigitated capacitors denoted $C_{c}$.}%
\label{cjphoto}%
\end{center}
\end{figure}

In order to mitigate some of the drawbacks of the MRJ, we developed a new JPC
design based on compact resonators known as CRJ. The circuit model of the CRJ,
shown in Fig. \ref{CJcirc}, consists of four equal capacitors denoted as $C$
and two pairs of linear inductors connected in series with the JRM whose total
inductance is $L_{a}$ and $L_{b}$ respectively. Using symmetry considerations
one can verify that this circuit has three eigenmodes. Two differential
eigenmodes which resonate at bare angular frequencies $\omega_{a}%
=1/\sqrt{\left(  L_{a}+L_{J}^{eff}\right)  C}$, $\omega_{b}=1/\sqrt{\left(
L_{b}+L_{J}^{eff}\right)  C}$, where $L_{J}^{eff}$ is the equivalent Josephson
inductance of the JRM biased at half a flux quantum, and an even eigenmode
which resonates at a lower bare angular frequency $\omega_{e}=1/\sqrt{\left(
L_{a}+L_{b}+L_{J}^{eff}\right)  C}$. Figure \ref{cjphoto} shows an optical
image of a typical compact JPC. The resonators of the device are made of Nb
deposited over sapphire substrate. They are fabricated using a standard
photolithography step and RIE etching. The JRM at the center of the device is
made of Aluminum. It is fabricated using e-beam lithography, and angle shadow
evaporation. As can be seen in the figure, the capacitance elements (including
the coupling capacitors) of the device are implemented using interdigitated
capacitors, whereas the inductive elements are realized using long narrow
superconducting lines. Unlike the microstrip resonator JPC, the compact
resonator JPC does not have higher harmonic resonances. The next closest
resonance of this structure resides above $4\omega_{a}$, therefore the pump
applied at $\omega_{a}+\omega_{b}$ can be considered stiff to a very good
approximation. Other advantages of this realization are: (1) small size, with
dimensions much smaller than the wavelengths corresponding to the resonance
frequencies, (2) no requirement of a definite ground plane, unlike the MRJ,
(3) greater flexibility in engineering the characteristic impedance of the
resonators higher or lower than 50 $\Omega$, (4) higher internal quality factor resonators than the microstrip design.
On the other hand, the main disadvantages of this design are: (1) the narrow
lines and the interdigitated capacitors (as well as the lines connecting them)
have parasitic capacitances and parasitic inductances associated with them,
therefore scaling these devices to match a certain frequency or certain
characteristic impedance requires using a microwave simulation tool, (2) there
is a limit to how big the capacitance can be using the interdigitated
configuration (values above 0.5 pF is difficult to achieve), therefore engineering characteristic impedances
below 30 $\Omega$ is not quite feasible with this design.%
%
%
\subsection{Shunted JPC (SJ)}%
%
%
\begin{figure}[h]
\begin{center}
\includegraphics[width=\columnwidth]{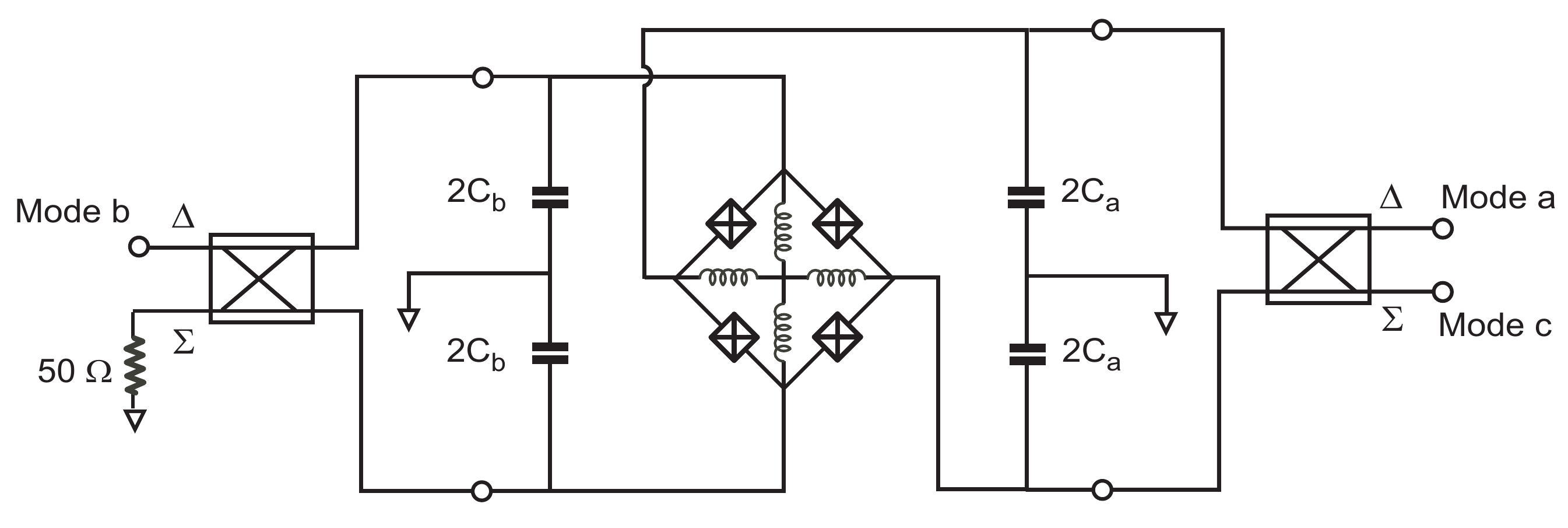}\\
\caption{Circuit model of the Shunted JPC (SJ).}%
\label{LJcirc}%
\end{center}
\end{figure}
In this subsection we discuss a third promising design called the capacitively
and inductively shunted JPC (SJ) which is still a work in progress in our lab.
In this version of the JPC, the capacitive elements are parallel plate
capacitors and the inductive elements are mainly Josephson junctions. A
schematic circuit model of the SJ is drawn in Fig. \ref{LJcirc}. It is
straightforward to show that the SJ model has two differential eigenmodes with
angular resonance frequencies $\omega_{a}=1/\sqrt{L_{J}^{^{\prime}}C_{a}}$,
$\omega_{b}=1/\sqrt{L_{J}^{^{\prime}}C_{b}}$, where $L_{J}^{^{\prime}}$
corresponds to the equivalent inductance of the JRM shunted by linear
inductors \cite{Roch}, as shown in Fig. \ref{LJcirc}. The main purpose of
these shunting inductors is to eliminate the hysteretic flux response of the
JRM and extend the frequency tunability of the device beyond the bandwidth
limit of the resonators. Such frequency tunability is achieved by varying the
flux threading the loop which, in turn, varies $L_{J}^{^{\prime}}$. Note that
the addition of these shunting inductors can be employed in other realizations
of the JPC also, such as the MRJ, as shown in Ref. \cite{Roch} and the CRJ. It
is important to emphasize, however, that the main difference between the SJ
and the CRJ or MRJ schemes is that the shunted JRM in the SJ design is the
only inductive element in the circuit that forms an integral part of the
resonators \textit{a} and \textit{b}. Thus, the larger lumped capacitors
employed in the SJ design play a crucial role in keeping the resonance
frequencies of the device below $10$ GHz.
\par
Similar to the Josephson bifurcation amplifier (JBA) implementation
\cite{VijayJBAreview}, the plate capacitors in the SJ design can be made of Nb
electrodes separated by a thin SiN dielectric layer. Using plate capacitors in
this realization has two advantages: (1) the plate capacitors can be made very
large, i.e. their capacitance can vary in the range $1-40$ pF, (2) they are easy to design as their capacitance scales linearly with the
electrode area. Furthermore, due to the lumped nature of the capacitive and
inductive elements in the SJ design and the fact that the capacitors can be
large, the SJ has three important advantages over the previous designs: (1)
the characteristic impedance of the resonators can be of the order of a few
ohms, which yields an improved coupling between the resonators and the JRM,
(2) due to the impedance mismatch between the characteristic impedance of the
resonators and the 50 $\Omega$ feedlines, the coupling capacitors are unnecessary to achieve low external Q
and the feedlines can be connected directly to the resonators, (3) the maximum
input power of the amplifier can be increased by increasing the critical
current of the JRM junctions while keeping the resonance frequencies fixed by
enlarging the capacitors.
%
%
\section{Experimental results}
%
%
The set of JPC parameters which can be directly measured in an experiment are: the
angular resonance frequencies of the resonators \textit{a} and \textit{b
}$\omega_{a}$, $\omega_{b}$, the inverse of residence times of photons at
resonance $\gamma_{a}$, $\gamma_{b}$, the participation ratios $p_{a}$,
$p_{b}$, the maximum input power which the device can handle with no applied
pump tone $P_{a}^{\max}$, $P_{b}^{\max}$, and the maximum measured gain at
vanishing input power $G_{0}^{\max}$.

One way to find $p_{a}$, $p_{b}$ is by measuring $\omega_{a}$, $\omega_{b}$ as
a function of applied magnetic flux threading the JRM loop. To establish this
relation, we model the resonators near resonance as an LC oscillator with
effective inductance $L_{a,b}$ and effective capacitance $C_{a,b}$. In this
model, the bare angular resonance frequencies of the device (with the
junctions) $\omega_{a}$, $\omega_{b}$, can be written as%

\begin{equation}
\omega_{a,b}\left(  \varphi\right)  =\frac{1}{\sqrt{C_{a,b}\left(
L_{a,b}+L_{J}\left(  \varphi\right)  \right)  }},
\end{equation}
where $L_{J}\left(  \varphi\right)  $ is the effective Josephson inductance of
the JRM given by%

\begin{equation}
L_{J}\left(  \varphi\right)  =\frac{L_{J}}{\cos\left(  \frac{\varphi}%
{4}\right)  }%
\end{equation}
with $\varphi=2\pi\Phi/\Phi_{0}$. By calculating the derivative of
$\omega_{a,b}\left(  \varphi\right)  $ with respect to the reduced flux
$\varphi$, one gets%

\begin{align}
\frac{1}{\omega_{a,b}}\frac{\mathrm{d}\omega_{a,b}}{\mathrm{d}\varphi}  &
=-\frac{1}{8}\tan\left(  \frac{\varphi}{4}\right)  \frac{L_{J}\left(
\varphi\right)  }{\left(  L_{a,b}+L_{J}\left(  \varphi\right)  \right)  },\\
&  =-\frac{1}{8}\tan\left(  \frac{\varphi}{4}\right)  p_{a,b}(\varphi).
\end{align}

Hence, at the device working point $\Phi=\Phi_{0}/2$ ($\varphi=\pi$),
$p_{a,b}$ reads%

\begin{equation}
p_{a,b}=-8\left.  \left(  \frac{1}{\omega_{a,b}}\frac{\mathrm{d}\omega_{a,b}%
}{\mathrm{d}\varphi}\right)  \right\vert _{\varphi=\pi}.
\end{equation}

Furthermore, using Eq. (\ref{p-min}) and the measured values $P_{a}^{\max}$,
$P_{b}^{\max}$, one can infer the Josephson energy $E_{J}^{a,b}$ which is
available for amplification%

\begin{equation}
E_{J}^{a,b}=p_{a,b}\frac{P_{a,b}^{\max}}{\gamma_{a,b}}.
\end{equation}

It is important to mention that, in our experiments, we find that this value
is lower by about one order of magnitude than the Josephson energy of the
junctions at the working point $E_{J}=I_{0}\varphi_{0}/\sqrt{2}$, where
$I_{0}$ is evaluated using dc resistance measurement of the junctions.

Using Eqs. (\ref{rho}) and (\ref{gain-formula}) for the case of maximum gain
$G_{0}^{\max}$ yields%

\begin{equation}
g_{3}^{2}\overline{n}_{c,\rho\rightarrow1}=\frac{\gamma_{a}\gamma_{b}}{4}%
\frac{\sqrt{G_{0}^{\max}}-1}{\sqrt{G_{0}^{\max}}+1}, \label{g3_sq_times_n_c}%
\end{equation}
which in the limit of high gains gives an upper bound on the product
$g_{3}^{2}\overline{n}_{c,\rho\rightarrow1}$%

\begin{equation}
g_{3}^{2}\overline{n}_{c,\rho\rightarrow1}\leq\frac{\gamma_{a}\gamma_{b}}{4}.
\end{equation}
Here $\overline{n}_{c,\rho\rightarrow1}$ is the number of pump photons in the
device at $G_{0}^{\max}$.%
\begin{figure}
[h]
\begin{center}
\includegraphics[width=\columnwidth]{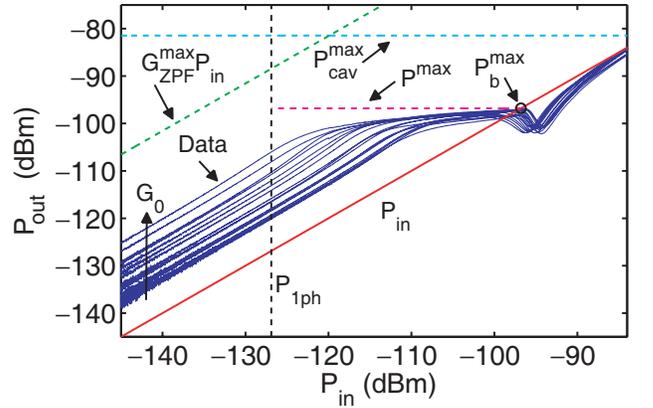}\\
\caption{(Color online). Output power $P_{\mathrm{out}}$ measurement of a
CRJ\ amplifier (device A) as a function of input power $P_{\mathrm{in}}$
measured at $\omega_{b}$. The data curves plotted in blue correspond to
different $G_{0}$ setpoints obtained for different pump powers. The red line
corresponds to $0$ dB (unity gain) where $P_{\mathrm{out}}=P_{\mathrm{in}}$.
The dashed black vertical line indicates the input power of $1$ photon at the
signal frequency per inverse dynamical bandwidth of the device at $G_{0}=20$
dB. The top horizontal line labelled $P_{\mathrm{cav}}^{\max}$ corresponds to
the maximum circulating power in the resonator cavity given by Eq.
(\ref{Pmaxcav}). The green line corresponds to an upper limit on the device
gain set by the saturation of the amplifier due to zero-point fluctuations
given by Eq. (\ref{Gmaxzpf}). The dashed magenta line is a theoretical
prediction for $P_{\mathrm{out}}^{\max}$, which corresponds to the maximum
circulating power in the device given by Eq. (\ref{DRstiff}). The measured and
calculated parameters of this device (A) are listed in table III.}%
\label{drcjfig}%
\end{center}
\end{figure}
In Figs. (\ref{drcjfig}), (\ref{drpsfig}), (\ref{drchfig}) we plot on
logarithmic scale the output power $P_{\mathrm{out}}$ of three different JPCs
with different characteristics as a function of input power $P_{\mathrm{in}}$.
For simplicity, we refer to the three devices as A, B and C respectively. The
parameters of the three devices are listed in table III. The data curves
plotted in blue are measured at resonance and satisfy the relation
\begin{equation}
P_{\mathrm{out}}=G\left(  P_{\mathrm{in}},G_{0}\right)  P_{\mathrm{in}},
\end{equation}
where $G\left(  P_{\mathrm{in}},G_{0}\right)  $ is the amplifier gain. This
depends on $P_{\mathrm{in}}$ and $G_{0}$, the device gain for $P_{\mathrm{in}%
}=0$ which is set by the applied pump power. In this measurement, we apply a
fixed pump power and vary $P_{\mathrm{in}}$ treating $G_{0}$ as a parameter.
In log units, the device gain translates into a vertical offset from the $0$
dB baseline (red line) which corresponds to $P_{\mathrm{out}}=P_{\mathrm{in}}$.

As expected, the devices maintain an almost constant gain $G_{0}$ as a
function of $P_{\mathrm{in}}$ before they saturate and their gain drops for
elevated input powers. However, as can be seen in Figs. (\ref{drcjfig}),
(\ref{drpsfig}), (\ref{drchfig}), the three devices exhibit qualitatively
different behaviors in the vicinity of their maximum input power, which
correspond to different saturation mechanisms taking place in the device as
will be discussed shortly. Note that the order in which the different results
are presented in this section does not depend on the specific implementation
of the device (see Sec. IV) but rather on the saturation mechanism involved in
each case.

In Fig. \ref{drcjfig}, device A exhibits almost a plateau in $P_{\mathrm{out}%
}$ as it reaches its maximum input power for different $G_{0}$ setpoints. This
result can be explained by assuming a stiff pump for which Eq. (\ref{P-Max})
applies. By employing $P_{b}^{\max}$, measured with no applied pump tone, we
plot the dashed magenta line labelled $P^{\max}$ which corresponds to
\begin{equation}
P_{\mathrm{out}}^{\max}=P_{b}^{\max}. \label{DRstiff}%
\end{equation}
The dashed black vertical line indicates the input power of $1$ photon at the
signal frequency per inverse dynamical bandwidth of the device at $G_{0}=20$ dB. In practice, as we discuss in Sec. VI, the usable region in the
parameter space of the device with respect to gain, bandwidth and maximum
input power lies within the boundaries of the fictitious triangle formed by
the magenta, red and black lines.%
\begin{figure}[h]
\begin{center}
\includegraphics[width=\columnwidth]{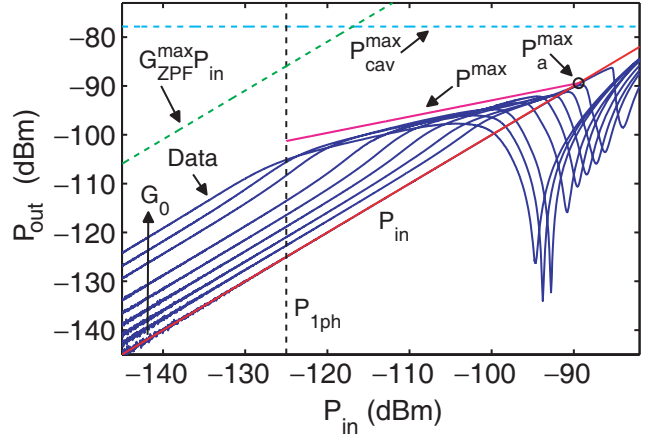}\\
\caption{(Color online). Output power $P_{\mathrm{out}}$ measurement of a
MRJ\ amplifier (device B) as a function of input power $P_{\mathrm{in}}$
measured at $\omega_{a}$. The data curves plotted in blue correspond to
different $G_{0}$ setpoints obtained for different pump powers. The red line
corresponds to $0$ dB (unity gain) where $P_{\mathrm{out}}=P_{\mathrm{in}}$.
The dashed black vertical line indicates the input power of $1$ photon at the
signal frequency per inverse dynamical bandwidth of the device at $G_{0}=20$
dB. The top horizontal line labelled $P_{\mathrm{cav}}^{\max}$ corresponds to
the maximum circulating power in the resonator cavity given by Eq.
(\ref{Pmaxcav}). The green line corresponds to an upper limit on the device
gain set by the saturation of the amplifier due to zero-point fluctuations
given by Eq. (\ref{Gmaxzpf}). The solid magenta line is a theoretical
prediction for $P_{\mathrm{in}}^{\max}$ of the device and the corresponding
$P_{\mathrm{out}}^{\max}$ due to pump depletion effect given by Eq.
(\ref{DRpumpDep}). The measured and calculated parameters of this device (B)
are listed in table III.}%
\label{drpsfig}%
\end{center}
\end{figure}
Furthermore, in Figs. (\ref{drcjfig}), (\ref{drpsfig}), (\ref{drchfig}) we
plot two fundamental limits on the maximum gain $G_{\mathrm{ZPF}}^{\max}$
(green line) which corresponds to saturation of the device due to amplified
zero-point fluctuations and the maximum circulating power $P_{\mathrm{cav}%
}^{\mathrm{\max}}$ (cyan line), given by Eq. (\ref{Gmaxzpf}) and Eq.
(\ref{Pmaxcav}) respectively.

The fact that these lines lie considerably above the experimental data in
Figs. (\ref{drcjfig}), (\ref{drpsfig}), (\ref{drchfig}), suggests that the
energy threshold, at which nonlinear effects in these devices become
significant, is much lower than the Josephson energy of the junctions, i.e.
$E_{J}^{a,b}\ll E_{J}$.

In contrast to Fig. \ref{drcjfig}, the data curves shown in Fig. \ref{drpsfig}
for device B, exhibit a gradual decrease in the gain in the vicinity of the
maximum input power which can be explained in terms of pump depletion effect
discussed in Sec. III. The maximum bound $P^{\max}$ indicated by the solid
magenta line corresponds to $P_{\mathrm{out}}^{\max}=G_{0}P_{\mathrm{in}%
}^{\max}$, where in this case $P_{\mathrm{in}}^{\max}$ satisfies the
inequality \ref{pumpDepletion1} and is given by
\begin{equation}
P_{\mathrm{in}}^{\max}=\frac{P_{a}^{\max}}{G_{0}^{3/2}}. \label{DRpumpDep}%
\end{equation}
\begin{figure}[h]
\begin{center}
\includegraphics[width=\columnwidth]{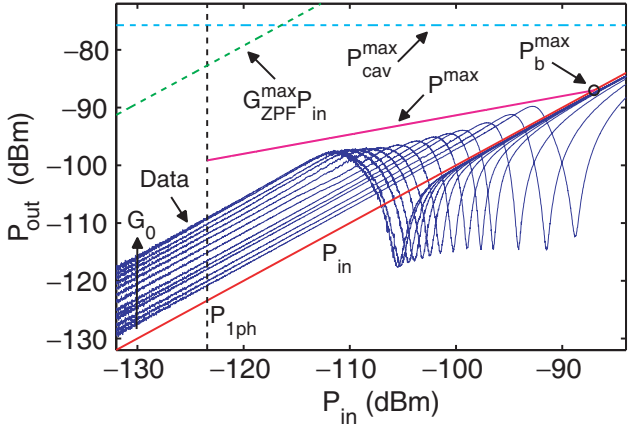}\\
\caption{(Color online). Output power $P_{\mathrm{out}}$ measurement of
another CRJ\ amplifier (device C, with different parameters from device A) as
a function of input power $P_{\mathrm{in}}$ measured at $\omega_{b}$. The data
curves plotted in blue corresponds to different $G_{0}$ setpoints obtained for
different pump powers. The data curves of this device exhibit abrupt drop in
the gain in the vicinity of the maximum input powers which suggests that the
device enters an unstable regime at elevated input powers. The red line
corresponds to $0$ dB (unity gain) where $P_{\mathrm{out}}=P_{\mathrm{in}}$.
The dashed black vertical line indicates the input power of $1$ photon at the
signal frequency per inverse dynamical bandwidth of the device at $G_{0}=20$
dB. The top horizontal line labelled $P_{\mathrm{cav}}^{\max}$ corresponds to
the maximum circulating power in the resonator cavity given by Eq.
(\ref{Pmaxcav}). The green line correspond to an upper limit on the device
gain set by the saturation of the amplifier due to zero-point fluctuations
given by Eq. (\ref{Gmaxzpf}). The solid magenta line is a theoretical
prediction for $P_{\mathrm{in}}^{\max}$ of the device and the corresponding
$P_{\mathrm{out}}^{\max}$ due to pump depletion effect given by Eq.
(\ref{DRpumpDep}). The measured and calculated parameters of this device (C)
are listed in table III.}%
\label{drchfig}%
\end{center}
\end{figure}
On the other hand the data curves shown in Fig. \ref{drchfig} for device C
exhibit an abrupt drop in the device gain in the vicinity of $P_{\mathrm{in}%
}^{\max}$ of the device, which indicates that the device enters an unstable
regime at elevated input powers. As can be seen in this case the solid magenta
line --- which satisfies $P_{\mathrm{out}}^{\max}=G_{0}P_{\mathrm{in}}^{\max}$,
where $P_{\mathrm{in}}^{\max}$ is given by Eq. (\ref{DRpumpDep}) --- lies above
the experimental data. This suggests that the maximum input power in this
sample, which displays a steeper power scaling than Eq. (\ref{DRpumpDep}), is
mainly limited by nonlinear effects arising from higher order terms in the
Hamiltonian of the system and cannot be attributed to a pump depletion effect
alone. It is worthwhile noting that a similar power scaling for the maximum
input power has been observed as well for an MRJ amplifier in Ref. \onlinecite{Jamp}.

To understand which properties are responsible for the different gain
behaviors exhibited by devices A, B and C, we point out a few important
distinctions in their design (respective parameters are listed in table III).
The data in Fig. \ref{drcjfig} (device A) and Fig. \ref{drchfig} (device C) is
measured on JPC devices realized using the CRJ configuration which yields, in
general, a stiff pump response as explained in Sec. IV (B). However, the main
two differences between devices A and C are: (1) device A has a narrower
bandwidth as compared to C (70 MHz vs. 142 MHz) and (2) the JRM junctions in A have a smaller $I_{0}$ compared to those in
C (2 $\mu$A vs. 4 $\mu$A). The relatively large bandwidth of device C leads to a larger dynamical
bandwidth 14 MHz at $G_{0}=16$ dB, as opposed to 10 MHz achieved in device A for the same gain, and also yields (with the larger
$I_{0}$ of device C) higher $P_{a,b}^{\max}$ values. However, the large
bandwidth translates into a lower $pQ$ product for C as compared to A, thus
making it more susceptible to parametric oscillation (at high gains or high
input powers) as implied by inequality (\ref{PQ-product}).

Device B, on the other hand, exhibits a pump depletion effect as shown in Fig.
\ref{drpsfig}. This can be attributed to its MRJ\ configuration, which, in
general, exhibits a less stiff pump response than the CRJ, due to the presence
of high order modes as explained in Sec. IV (A). Furthermore, as opposed to
the MRJ amplifier in Ref. \onlinecite{Jamp} with an idler frequency of 6.4 GHz, device B has a higher idler frequency of 15 GHz which leads to a higher $pQ$ product.
\begin{table}[htb]%
\begin{tabular}[c]{|c|c|c|c|}
\hline
\ \textrm{Parameter }$\backslash$\ \textrm{Device} & \ \textrm{A} & \ \textrm{B} & \ \textrm{C}\\
\hline\hline
\ \textrm{Design} & \ \textrm{CRJ} & \ \textrm{MRJ} & \ \textrm{CRJ}\\
\hline
$\omega_{a}/2\pi$ \ \textrm{(GHz)} & $6.576$ & $8.436$ & $7.051$\\
\hline
$\omega_{b}/2\pi$ \ \textrm{(GHz)} & $6.873$ & $15.087$ & $7.673$\\
\hline
$\omega_{3}/2\pi$ \ \textrm{(GHz)} & $13.449$ & $23.523$ & $14.724$\\
\hline
$\gamma_{a}/2\pi$ \ \textrm{(MHz)} & $69$ & $116$ & $79$\\
\hline
$\gamma_{b}/2\pi$ \ \textrm{(MHz)} & $71$ & $250\pm25$ & $142$\\
\hline
$Q_{a},Q_{b}$ & $94,96$ & $73,60$ & $89,54$\\
\hline
$p_{a},p_{b}$ & $0.02$ & $0.03,0.05$ & $0.03$\\
\hline
$p_{a}p_{b}Q_{a}Q_{b}$ & $8.1$ & $6.6$ & $4.3$\\
\hline
$I_{0}$ $(\mu A)$ & $2$ & $3$ & $4$\\
\hline
$P_{\;\mathrm{cav}}^{\;\mathrm{max}}$ \ \textrm{(dBm)} & $-82$ & $-77$ & $-76$\\
\hline
$P_{a,b}^{\;\mathrm{max}}$ \ \textrm{(dBm)} & $-97$ & $-89$ & $-87$\\
\hline
$P_{\;\mathrm{1ph}}$ \ \textrm{(dBm)} & $-127$ & $-125$ & $-123$\\
\hline
$G_{\;\mathrm{ZPF}}^{\;\mathrm{max}}$ \ \textrm{(dB)} & $38$ & $39$ & $40$\\
\hline
$G_{0}^{\;\mathrm{max}}$ \ \textrm{(dB)} & $22$ & $20$ & $16$\\
\hline
$g_{3}\bar{n}_{c,\rho\rightarrow1}^{1/2}/2\pi$ \ \textrm{(MHz)} & $33$ & $77$ & $45$\\
\hline
\end{tabular}
\label{Table3}
\caption{Parameters of JPCs A, B and C. Precision is last significant digit unless indicated otherwise.}%
\end{table}
%
%
\section{Requirements for qubit readout}
%
%
One of the leading architectures which is used to manipulate and readout the
state of superconducting qubits such as transmons and fluxoniums
\cite{TransmonThy,fluxonium} is circuit Quantum Electrodynamics (cQED). In
such a system a quantum non-demolition measurement of the qubit state can be
performed using dispersive readout in which the frequencies of the qubit and
the cavity are detuned. As a result, the qubit and the cavity interact via
exchanging virtual microwave photons \cite{RSL1} and the qubit state gets
encoded in the output microwave field of the cavity. However, since the energy
of microwave photons is very small, the detection of single photons is
difficult especially considering the fact that state-of-the-art cryogenic
amplifiers (i.e. high electron mobility transistor (HEMT) \cite{HEMT})
following the cQED setup add noise to the input signal, equivalent to about
$20-40$ photons at the signal frequency. Therefore, adding a quantum-limited
amplifier in series between the cQED sample and the HEMT amplifier can
substantially decrease the noise temperature of the system and enable
real-time tracking of the qubit state \cite{QuantumJumps,QubitJPC}. The
desired requirements of a Josephson parametric amplifier for such
high-fidelity qubit readout can be summarized as follows:
\begin{itemize}
\item A center frequency in the range $5-12$ GHz which is widely used in readout cavities of superconducting qubits.

\item A large power gain on the order of $20$ dB in order to beat the noise of the following amplifier, i.e. the HEMT.

\item A minimum added noise to the signal, equivalent to a half input photon at the signal frequency $T_{\mathrm{N}}=\hbar\omega_{a,b}/2k_{B}$ when
operated in the phase preserving mode \cite{Caves}.

\item A large dynamical bandwidth of the order of $10$ MHz, which corresponds to a signal processing time of less than $100$ ns and matches the bandwidth of most readout cavities.

\item A maximum input power of a few photons per inverse dynamical bandwidth of the device at the highest gain. Such requirement is essential in quantum non-demolition readout schemes which employ of the order of a photon on average \cite{fluxonium}.

\item A tunable bandwidth of more than 100 MHz so that the center frequency of the amplifier can match the frequency of the readout signal. Recently, Roch \textit{et al.} \cite{Roch} have achieved a tunable bandwidth of more than 500 MHz in a MRJ device by shunting the Josephson junctions of the JRM with linear inductors realized using superconducting wires [see Fig. \ref{LJcirc}]. Similar results were obtained by our group in a MRJ device by utilizing large Josephson junctions instead of superconducting wires \cite{QubitJPC}.

\item Minimal out-of-band back-action to avoid qubit relaxation.
\end{itemize}
In table IV we enlist the parameters achievable with a JPC which show its
viability as a low-noise preamplifier for qubit
measurements.
\begin{table}[htb]%
\begin{tabular}[c]{|c|c|c|}
\hline
\ \textrm{Property} & \ \textrm{Requirement} & \ \textrm{Achieved to date \cite{Jamp}}\\
\hline\hline
$\omega_{a,b}/2\pi$ & 5 - 12\ \textrm{GHz} & 6.4 \& 8.1 \ \textrm{GHz}\\
\hline
$T_{\;\mathrm{N}}$ $@$ 8\ \textrm{GHz} & 0.2\ \textrm{K} & 0.4\ \textrm{K}\\
\hline
$G$ & $20$\ \textrm{dB} & $21$\ \textrm{dB}\\
\hline
$B$ & 10\ \textrm{MHz} & 11\ \textrm{MHz}\\
\hline
\ \textrm{Tunable BW} & 100\ \textrm{MHz} & 60\ \textrm{MHz}\\
\hline
$P_{\;\mathrm{max}}$ & $1$ photon & 3 photons $@$ 20 dB\\
\hline
OB back-action & Negligible & None measurable\\
\hline
\end{tabular}
\label{Table4}
\caption{Preamplifier requirements and JPC merits achieved to date (OB=out-of-band).}%
\end{table}
\par
The last question which we would like to address in this paper is whether
there exists a set of technologically feasible parameters for which the JPC
can be optimized with respect to dynamic range while maintaining a gain of
$20$ dB and a reasonable dynamical bandwidth of more than 2 MHz. In order to provide a quantitative answer we choose a signal frequency of
12 GHz, which is a good choice for readout frequency for qubits as it is higher
than most qubit frequencies. We set an ambitious goal for the processing
capability of the JPC of about $100$ input photons at the signal frequency
with gain $20$ dB.%
\begin{figure}[h]
\begin{center}
\includegraphics[width=\columnwidth]{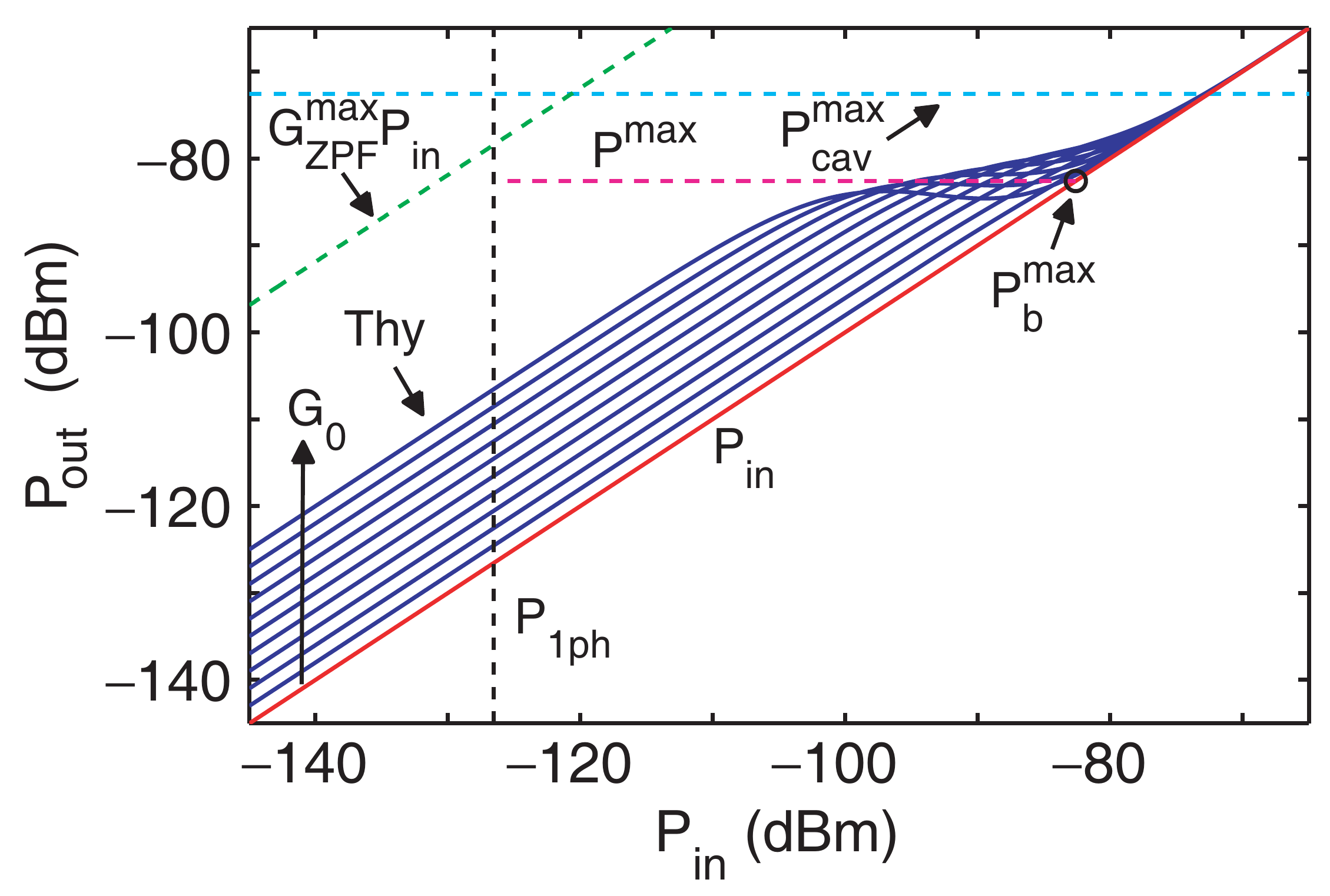}\\
\caption{(Color online). An optimized JPC response drawn in blue for large
maximum input power in excess of $100$ input photons at $12$
$\operatorname{GHz}$ per inverse dynamical bandwidth of the device at $20$ dB. The
definition of the other lines shown in the figure is similar to Fig.
(\ref{drcjfig}). The parameters used in the calculation are: $\omega_{a}%
/2\pi=11\operatorname{GHz}$, $\omega_{b}/2\pi=12\operatorname{GHz}$,
$\omega_{c}/2\pi=23\operatorname{GHz}$, $Z_{a}=36\operatorname{\Omega }$,
$Z_{b}=33\operatorname{\Omega }$, $L_{a}=0.51$ n$\operatorname{H}$,
$L_{b}=0.42$ n$\operatorname{H}$, $C_{a}=C_{b}=0.4\operatorname{pF}$,
$C_{C_{a}}=31$ f$\operatorname{F}$, $C_{C_{b}}=28$ f$\operatorname{F}$,
$\gamma_{a}/2\pi=\gamma_{b}/2\pi=44.2\operatorname{MHz}$, $\gamma_{c}%
/2\pi=3\operatorname{GHz}$, $Q_{a}=249$, $Q_{b}=271$, $Q_{c}=8$, $p_{a}%
=0.028$, $p_{b}=0.034$, $p_{c}=0.02$, $I_{0}=30\operatorname{\mu A}$,
$E_{J}/\sqrt{2}=490\operatorname{K}$, $E_{J}^{a,b}=49\operatorname{K}$,
$P_{\mathrm{1ph}}=-126$ dBm, $G_{\mathrm{ZPF}}^{\max}=48$ dB, $P_{\mathrm{cav}%
}^{\mathrm{\max}}=P_{b}^{\mathrm{\max}}=-86.3$ dBm and $g_{3}/2\pi
=0.6\operatorname{MHz}$. }%
\label{DRprospectsfig}%
\end{center}
\end{figure}
To that end, we choose to perform the optimization process for the CRJ
configuration which inherently yields large $\gamma_{c}$ values and also
allows variation of the external quality factor of the resonators more easily
than the SJ scheme. We also choose a resonance frequency for mode \textit{a}
of $\omega_{a}/2\pi=$ 11 GHz, a relatively high critical current $I_{0}=$ 30 $\mu$A and limit ourselves to capacitance values below or equal to 0.4 pF. The advantage of working with large $I_{0}$ for the purpose of large
dynamic range is that it increases $E_{J}$ and lowers $g_{3}$. However, such
large $I_{0}$ yields very low $L_{J}^{eff}=\sqrt{2}\varphi_{0}/I_{0}=15$ pH which requires coupling to relatively low impedance resonators while
maintaining participation ratios of a few percent. The next challenge in the
optimization is to increase the $pQ$ product of the device which promotes high
gains by increasing the quality factor of the resonators. Nevertheless, care
must be taken not to increase the quality factors beyond what is strictly
necessary for two reasons (i) high Q resonators limit the dynamical bandwidth
of the device as can be seen in Eq. (\ref{B}), (ii) high Q resonators increase
the pump depletion effect and, in turn, lower the dynamic range of the device.
In Fig. \ref{DRprospectsfig} we plot the calculated response of such an
optimized JPC which takes into account the above considerations and
limitations. As can be seen in the figure the optimized device exhibits, for
the chosen set of parameters, a maximum input power of about $100$ photons at
the signal frequency per inverse dynamical bandwidth of the device $B/2\pi =$ 4.4 MHz at $20$ dB of gain. The device parameters which are used in the calculation
are listed in the figure caption. It is important to note that in the
calculation of the expected response, which is indicated by the blue curves
for different values of $G_{0}$, we assumed an available Josephson energy $10$
times smaller than $E_{J}/\sqrt{2}$ of the junctions, in agreement with
experimental conditions. Finally, we verify that the set of parameters of the
optimized device satisfy the inequalities $\bar{n}_{c}^{\max}=3.7\cdot10^{3}>\bar{n}_{c}^{\mathrm{po}}=1.3\cdot10^{3}> \bar{n}_{c}^{20\mathrm{dB}}=10^{3}$.
%
%
\section{Conclusion}
\label{conclusion}
%
%
We have addressed in this paper a new type of quantum signal processing device
based on Josephson tunnel junctions. In contrast with the devices based on
SQUIDS and driven non-linear Josephson oscillators, it performs a fully
non-degenerate three-wave mixing in which the modes of the signal, pump and
idler are separate both spatially and temporally. The heart of the device
consists of a ring modulator constructed from four Josephson junctions
arranged in a loop. Both quantum-limited amplification and noiseless frequency
conversion are possible with this device, and the characteristics of these
analog signal processing operations are entirely calculable analytically. We
have established the limitations preventing the simultaneous maximization of
photon number gain, bandwidth and dynamic range. Nevertheless, we have shown
that a device satisfying all the requirements of superconducting qubit readout
is realizable with present day technology.
%
%
%
%
\begin{acknowledgements}
Discussions with Flavius Schackert, Michael Hatridge, Nicolas Bergeal,
Benjamin Huard and Ananda Roy are gratefully acknowledged. The assistance of
Michael Power and Luigi Frunzio in the fabrication process is highly
appreciated. This work was supported by Yale University, NSF, IARPA, ARO and
College de France.
\end{acknowledgements}
%
%
%
%
\appendix
%
%
\section*{Appendix: Quantum signals propagating along a transmission line and input-output formalism}
\label{appendix Quantum signals}
\begin{figure}[h]
\begin{center}
\includegraphics[width=\columnwidth]{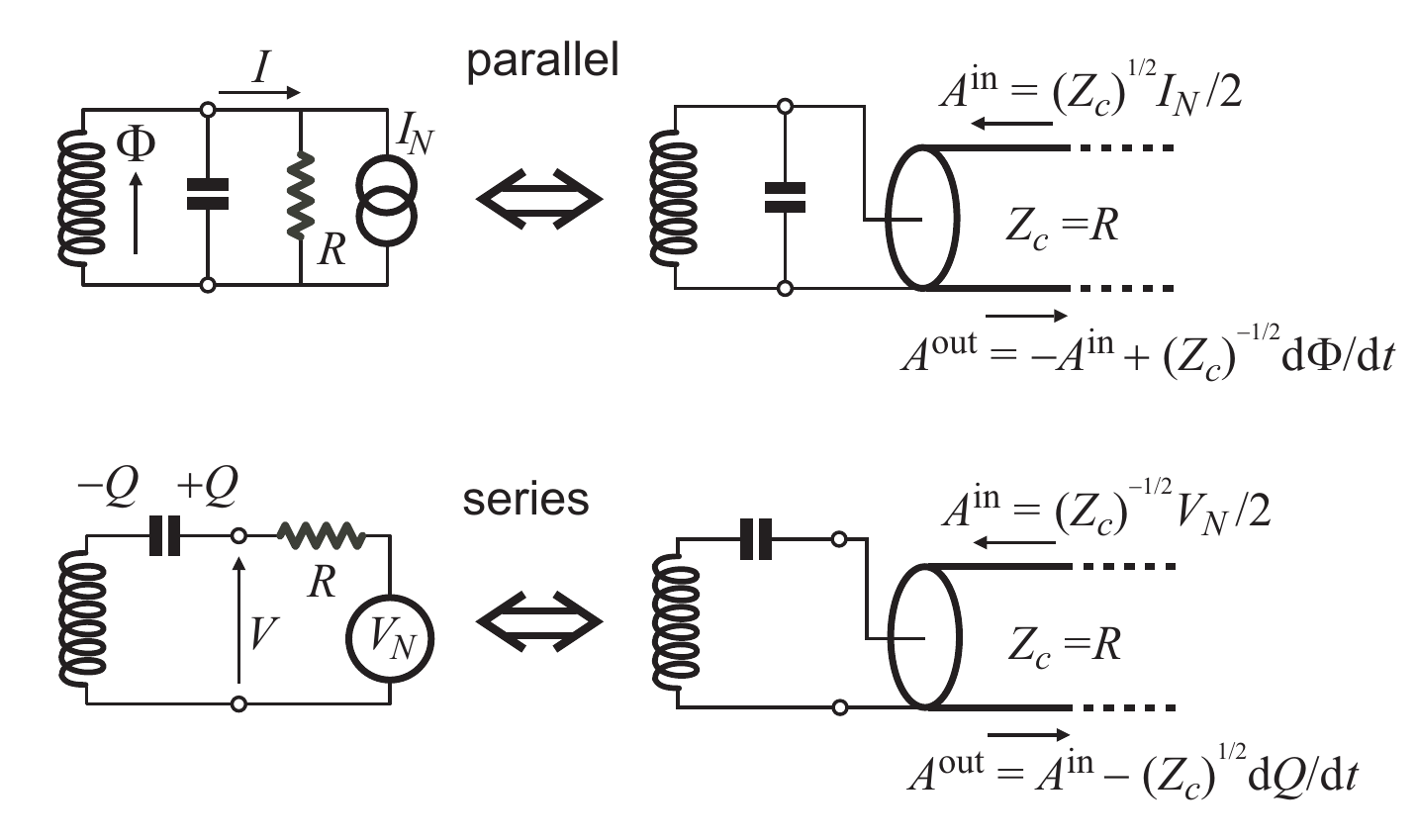}\\
\caption{The damping of a circuit by a resistance $R$ can take place in a
parallel or series way, depending on whether the resistance is placed across a
branch or in series with it. The Nyquist model represents the resistance by a
transmission line with characteristic impedance $Z_{c}=R$. The noise source
associated with the resistance (fluctuation-dissipation theorem) is a parallel
current source in the parallel case and a series voltage source in the series
case. The noise source is replaced in the Nyquist model by incoming thermal
radiation whose amplitude $A^{\mathrm{in}}$ is the square root of the power
flux of the radiation ($A^{\mathrm{in}}$ should not be associated to a vector
potential and is rather like the square root of the length of the Poynting
vector).}%
\label{Nyquist-model}
\end{center}
\end{figure}

%
%
This appendix treats quantum-mechanically the damping of a circuit by a
resistance modelled as a semi-infinite transmission line, as shown in Fig.
\ref{Nyquist-model}. It borrows heavily from the book by Gardiner and Zoller
\cite{QuantumNoise} but uses slightly different notations that are adapted to
the specificities of our Josephson circuits. We first describe an infinite
transmission line extending from $x=-\infty$ to $x=+\infty$. Later, we will
cut the line at $x=0$ and replace the left portion by two terminals of the circuit.%
%
%
\subsection*{Infinite transmission line}
%
%
The capacitance and inductance per unit length of the line are $C_{\ell}$ and
$L_{\ell}$, respectively. The equations obeyed by the current $I$ along and
the voltage $V$ across the line are
\begin{align}
-\frac{\partial}{\partial x}V\left(  x,t\right)   &  =L_{\ell}\frac{\partial
}{\partial t}I\left(  x,t\right)  ,\label{propagation_eq_1}\\
-\frac{\partial}{\partial x}I\left(  x,t\right)   &  =C_{\ell}\frac{\partial
}{\partial t}V\left(  x,t\right)  , \label{propagation_eq_2}%
\end{align}
in which, for the moment, we treat the fields classically. The characteristic
impedance and propagation velocity are given by
\begin{align}
Z_{c}  &  =\sqrt{\frac{L_{\ell}}{C_{\ell}}},\\
v_{p}  &  =\sqrt{\frac{1}{L_{\ell}C_{\ell}}}.
\end{align}
In order to solve Eqs. (\ref{propagation_eq_1}) and (\ref{propagation_eq_2}),
we introduce two new fields: the left-moving and right-moving wave
amplitudes,
\begin{align}
A^{\rightarrow}\left(  x,t\right)   &  =\frac{1}{2}\left[  \frac{1}%
{\sqrt{Z_{c}}}V\left(  x,t\right)  +\sqrt{Z_{c}}I\left(  x,t\right)  \right]
,\\
A^{\leftarrow}\left(  x,t\right)   &  =\frac{1}{2}\left[  \frac{1}{\sqrt
{Z_{c}}}V\left(  x,t\right)  -\sqrt{Z_{c}}I\left(  x,t\right)  \right]  ,
\end{align}
which have the advantage of treating currents and voltage on the same footing
(note that these amplitudes are not directly related to the vector potential).
The dimension of these fields is [watt]$^{1/2}$ and they are normalized such
that the total power $P$ traversing, in the forward direction, a section of
the line at position $x$ and time $t$ is given by
\begin{equation}
P\left(  x,t\right)  =\left[  A^{\rightarrow}\left(  x,t\right)  \right]
^{2}-\left[  A^{\leftarrow}\left(  x,t\right)  \right]  ^{2}. \label{poynting}%
\end{equation}
The quantity $P$ here plays the role of the Poynting vector in full 3D
electrodynamics. Each of the terms at the right hand side of the last equation
is thus the separate contribution of the corresponding wave to the total power flow.

When solving Eqs. (\ref{propagation_eq_1}-\ref{propagation_eq_2}), we find
\begin{equation}
\frac{\partial}{\partial x}A^{\rightleftarrows}\left(  x,t\right)  =\mp
\frac{1}{v_{p}}\frac{\partial}{\partial t}A^{\rightleftarrows}\left(
x,t\right)  . \label{Eq._of_motion}%
\end{equation}
This relation means that $A^{\rightleftarrows}$ does not depend separately on
$x$ or $t$ but a combination of both and thus:
\begin{align}
A^{\rightarrow}\left(  x,t\right)   &  =A^{\rightarrow}\left(  x=0,t-\frac
{x}{v_{p}}\right)  =A^{\rightarrow}\left(  x-v_{p}t,t=0\right)  ,\nonumber\\
A^{\leftarrow}\left(  x,t\right)   &  =A^{\leftarrow}\left(  x=0,t+\frac
{x}{v_{p}}\right)  =A^{\leftarrow}\left(  x+v_{p}t,t=0\right)  .\nonumber\\
&
\end{align}
The properties of the wave amplitude can be summarized by writing
\begin{align}
A^{\rightleftarrows}\left(  x,t\right)   &  =A_{0}^{\rightleftarrows}\left(
\tau\right)  ,\\
\tau &  =t+\frac{\varepsilon^{\rightleftarrows}}{v_{p}}x,\\
\varepsilon^{\rightleftarrows}  &  =\mp1.
\end{align}
Note that the detailed definition of the retardation $\tau$ depends on the
wave direction. We now turn to the energy density $U\left(  x,t\right)  $,
related to $P$ by the local energy conservation law
\begin{equation}
\frac{\partial U}{\partial t}=-\frac{\partial P}{\partial x}.
\end{equation}
Combining Eqs. (\ref{poynting}) and (\ref{Eq._of_motion}), we get
\begin{align}
&  \frac{\partial U\left(  x,t\right)  }{\partial t}\nonumber\\
&  =\frac{2}{v_{p}}\left[  A^{\rightarrow}\left(  x,t\right)  \frac{\partial
}{\partial t}A^{\rightarrow}\left(  x,t\right)  +A^{\leftarrow}\left(
x,t\right)  \frac{\partial}{\partial t}A^{\rightarrow}\left(  x,t\right)
\right]  ,\nonumber\\
&  =\frac{1}{v_{p}}\frac{\partial}{\partial t}\left\{  \left[  A^{\rightarrow
}\left(  x,t\right)  \right]  ^{2}+\left[  A^{\leftarrow}\left(  x,t\right)
\right]  ^{2}\right\}  .
\end{align}
The total energy of the line at time $t$ is, thus \cite{SumofEMenergy},
\begin{equation}
H=\frac{1}{v_{p}}\int_{-\infty}^{+\infty}\left\{  \left[  A^{\rightarrow
}\left(  x,t\right)  \right]  ^{2}+\left[  A^{\leftarrow}\left(  x,t\right)
\right]  ^{2}\right\}  \mathrm{d}x. \label{HinA}%
\end{equation}
When $H$ in Eq. (\ref{HinA}) is considered as a functional of dynamical field
variables $A^{\rightarrow}$ and $A^{\leftarrow}$, the equation of motion Eq.
(\ref{Eq._of_motion}) can be recovered from Hamilton's equation of motion as
\begin{equation}
\frac{\partial}{\partial t}A^{\rightleftarrows}\left(  x,t\right)  =-\left\{
H,A^{\rightleftarrows}\left(  x,t\right)  \right\}  _{P.B.},
\end{equation}
on imposing the Poisson bracket
\begin{align}
&  \left\{  A^{\rightleftarrows}\left(  x_{1},t_{1}\right)
,A^{\rightleftarrows}\left(  x_{2},t_{2}\right)  \right\}  _{P.B.}
  =\frac{1}{2}\frac{\partial}{\partial\left(  \tau_{1}-\tau_{2}\right)
}\delta\left(  \tau_{1}-\tau_{2}\right). \label{PBinA2}%
\end{align}
Therefore, from the classical-quantum correspondence involving the replacement
of Poisson brackets by commutators, we find that the quantum operator version
$\hat{A}^{\rightleftarrows}$ of the fields satisfy the commutation relation
\begin{equation}
\left[  \hat{A}^{\rightleftarrows}\left(  x_{1},t_{1}\right)  ,\hat
{A}^{\rightleftarrows}\left(  x_{2},t_{2}\right)  \right]  =\frac{i\hbar}%
{2}\frac{\partial}{\partial\left(  \tau_{1}-\tau_{2}\right)  }\delta\left(
\tau_{1}-\tau_{2}\right)  ,
\end{equation}
which is analogous to the commutation relation between the electric and
magnetic field in 3-D quantum electrodynamics. Note that the fields are
Hermitian at this stage. Introducing the Fourier transform,
\begin{equation}
\hat{A}^{\rightleftarrows}\left[  \omega\right]  =\frac{1}{\sqrt{2\pi}}%
\int_{-\infty}^{+\infty}\hat{A}^{\rightleftarrows}\left(  x=0,\tau\right)
e^{i\omega\tau}\mathrm{d}\tau,
\end{equation}
where the Fourier components (which are now non-hermitian operators) satisfy%

\begin{equation}
\hat{A}^{\rightleftarrows}\left[  \omega\right]  ^{\dag}=A^{\rightleftarrows
}\left[  -\omega\right]  ,
\end{equation}
we can also write the Hamiltonian as
\begin{equation}
\sum_{\sigma=\rightleftarrows}\int_{-\infty}^{+\infty}\hat{A}^{\sigma}\left[
\omega\right]  \hat{A}^{\sigma}\left[  -\omega\right]  \mathrm{d}\omega.
\end{equation}
The field operators in the frequency domain satisfy
\begin{equation}
\left[  \hat{A}^{\rightleftarrows}\left[  \omega_{1}\right]  ,\hat
{A}^{\rightleftarrows}\left[  \omega_{2}\right]  \right]  =\frac{\hbar}%
{4}\left(  \omega_{1}-\omega_{2}\right)  \delta\left(  \omega_{1}+\omega
_{2}\right)  .
\end{equation}
We now introduce the usual quantum field annihilation operators
\begin{align}
a^{\rightarrow}\left[  \omega\right]   &  =\frac{\hat{A}^{\rightarrow}\left[
\omega\right]  }{\sqrt{\hbar\left\vert \omega\right\vert /2}}=a^{\rightarrow
}\left[  -\omega\right]  ^{\dagger},\\
a^{\leftarrow}\left[  \omega\right]   &  =\frac{\hat{A}^{\leftarrow}\left[
\omega\right]  }{\sqrt{\hbar\left\vert \omega\right\vert /2}}=a^{\leftarrow
}\left[  -\omega\right]  ^{\dagger}.
\end{align}
They satisfy the commutation relations
\begin{equation}
\left[  a^{\rightleftarrows}\left[  \omega_{1}\right]  ,a^{\rightleftarrows
}\left[  \omega_{2}\right]  \right]  =\mathrm{sgn}\left(  \frac{\omega
_{1}-\omega_{2}}{2}\right)  \delta\left(  \omega_{1}+\omega_{2}\right)
.\label{bosonic_com}%
\end{equation}
It is useful to note that since
\begin{equation}
a^{\rightleftarrows}\left[  \omega\right]  =a^{\rightleftarrows}\left[
-\omega\right]  ^{\dag},
\end{equation}
Eq. (\ref{bosonic_com}) exhaustively describes all possible commutator cases.

In the thermal state of the line, at arbitrary temperature (including $T=0$),%

\begin{equation}
\left\langle a^{\rightleftarrows}\left[  \omega_{1}\right]
a^{\rightleftarrows}\left[  \omega_{2}\right]  \right\rangle
=S_{a^{\rightleftarrows}a^{\rightleftarrows}}\left[  \frac{\omega_{1}%
-\omega_{2}}{2}\right]  \delta\left(  \omega_{1}+\omega_{2}\right)  ,
\end{equation}
where%

\begin{equation}
S_{a^{\rightleftarrows}a^{\rightleftarrows}}\left[  \omega\right]
=\mathrm{sgn}\left(  \omega\right)  N_{T}\left(  \omega\right)  .
\end{equation}
When $\omega$ is strictly positive $N_{T}\left(  \omega\right)  $ is the
number of available photons per unit bandwidth per unit time travelling on the
line in a given direction around frequency $\omega$%

\begin{align}
N_{T}\left(  \omega\right)   &  =\frac{1}{\exp\left(  \frac{\hbar\omega}%
{k_{B}T}\right)  -1}\\
&  =\frac{1}{2}\left[  \coth\left(  \frac{\hbar\omega}{2k_{B}T}\right)
-1\right]  .
\end{align}
Negative frequencies $\omega$ correspond to the possibility of emitting
photons into the line%

\begin{equation}
N_{T}\left(  -\left\vert \omega\right\vert \right)  =-N_{T}\left(  \left\vert
\omega\right\vert \right)  -1.
\end{equation}
The Bose-Einstein expression $N_{T}\left(  \omega\right)  $ is expected from
the Hamiltonian of the line, which reads, with the $a$ operators,
\begin{equation}
H=\frac{\hbar}{2}\sum_{\sigma=\rightleftarrows}\int_{-\infty}^{+\infty
}\left\vert \omega\right\vert a^{\sigma}\left[  \omega\right]  a^{\sigma
}\left[  -\omega\right]  \mathrm{d}\omega.
\end{equation}
We can now give the expression for the anticommutator of the fields%

\begin{align}
&  \left\langle \left\{  a^{\rightleftarrows}\left[  \omega_{1}\right]
,a^{\rightleftarrows}\left[  \omega_{2}\right]  \right\}  \right\rangle
_{T}=2\mathcal{N}_{T}\left[  \frac{\omega_{1}-\omega_{2}}{2}\right]
\delta\left(  \omega_{1}+\omega_{2}\right) \nonumber\\
&  =\mathrm{sgn}\left(  \frac{\omega_{1}-\omega_{2}}{2}\right)  \coth\left(
\frac{\hbar\left(  \omega_{1}-\omega_{2}\right)  }{4k_{B}T}\right)
\delta\left(  \omega_{1}+\omega_{2}\right)  . \label{bos-anticom6}%
\end{align}
Equation (\ref{Na_in_first}) with no external drive is identical to Eq.
(\ref{bos-anticom6})%

\begin{align}
\mathcal{N}_{T}\left[  \omega\right]   &  =\frac{\mathrm{sgn}\left(
\omega\right)  }{2}\coth\left(  \frac{\hbar\omega}{2k_{B}T}\right)  \\
&  =\mathrm{sgn}\left(  \omega\right)  \left[  N_{T}\left(  \left\vert
\omega\right\vert \right)  +\frac{1}{2}\right]  .
\end{align}
We now introduce the forward-propagating and backward-propagating voltage and
current amplitudes obeying
\begin{align}
V^{\rightarrow}\left(  x,t\right)   &  =\sqrt{Z_{c}}A^{\rightarrow}\left(
x,t\right)  ,\\
V^{\leftarrow}\left(  x,t\right)   &  =\sqrt{Z_{c}}A^{\leftarrow}\left(
x,t\right)  ,
\end{align}%
\begin{align}
I^{\rightarrow}\left(  x,t\right)   &  =V^{\rightarrow}\left(  x,t\right)
/Z_{c},\\
I^{\leftarrow}\left(  x,t\right)   &  =V^{\leftarrow}\left(  x,t\right)
/Z_{c}.
\end{align}
Quantum-mechanically, the voltage and current amplitudes become hermitian
operators
\begin{align}
V^{\rightleftarrows}\left(  x,t\right)   &  \rightarrow\hat{V}%
^{\rightleftarrows}\left(  x,t\right)  ,\\
I^{\rightleftarrows}\left(  x,t\right)   &  \rightarrow\hat{I}%
^{\rightleftarrows}\left(  x,t\right)  .
\end{align}
These operators, in turn, can be expressed in terms of field annihilation
operators as
\begin{align}
\hat{V}^{\rightleftarrows}\left(  x,t\right)   &  =\sqrt{\frac{\hbar Z_{c}%
}{4\pi}}\int_{-\infty}^{+\infty}\mathrm{d}\omega\sqrt{\left\vert
\omega\right\vert }\hat{a}^{\rightleftarrows}\left[  \omega\right]
e^{-i\omega\left(  t\,\mp\,x/v_{p}\right)  },\\
\hat{I}^{\rightleftarrows}\left(  x,t\right)   &  =\sqrt{\frac{\hbar}{4\pi
Z_{c}}}\int_{-\infty}^{+\infty}\mathrm{d}\omega\sqrt{\left\vert \omega
\right\vert }\hat{a}^{\rightleftarrows}\left[  \omega\right]  e^{-i\omega
\left(  t\,\mp\,x/v_{p}\right)  }.
\end{align}
All physical operators can be deduced from these primary expressions. For
instance, the transmission line charge operator, describing the charge in the
line brought from one end to the position $x$, is
\begin{equation}
\hat{Q}^{\rightleftarrows}\left(  x,t\right)  =i\sqrt{\frac{\hbar}{4\pi Z_{c}%
}}\int_{-\infty}^{+\infty}\frac{\mathrm{d}\omega\sqrt{\left\vert
\omega\right\vert }}{\omega}\hat{a}^{\rightleftarrows}\left[  \omega\right]
e^{-i\omega\left(  t\,\mp\,x/v_{p}\right)  }.
\end{equation}
%
%
\subsection*{Nyquist model of resistance: semi-infinite transmission line}
%
%
We now are in a position to deal with the semi-infinite line extending from
$x=0$ to $x=\infty$, whose terminals at $x=0$ models a resistance $R=Z_{c}$
[see Fig. \ref{Nyquist-model}]. In that half-line, the left- and right-moving
propagating waves are no longer independent. We will now refer to the wave
amplitude $A^{\leftarrow}\left(  x=0,t\right)  $ as $A^{\mathrm{in}}\left(
t\right)  $ and $A^{\rightarrow}\left(  x=0,t\right)  $ as $A^{\mathrm{out}%
}\left(  t\right)  $. The quantum-mechanical voltage across the terminal of
the resistance and the current flowing into it satisfy the operator relations
\begin{align}
\hat{V}\left(  t\right)   &  =\hat{V}^{\mathrm{out}}\left(  t\right)  +\hat
{V}^{\mathrm{in}}\left(  t\right)  ,\\
\hat{I}\left(  t\right)   &  =\hat{I}^{\mathrm{out}}\left(  t\right)  -\hat
{I}^{\mathrm{in}}\left(  t\right)  .
\end{align}
These relations can be seen either as continuity equations at the interface
between the damped circuit and the resistance/line, or as boundary conditions
linking the semi-infinite line quantum fields $\hat{A}^{\mathrm{in}}\left(
t\right)  $ and $\hat{A}^{\mathrm{out}}\left(  t\right)  $. From the
transmission line relations,
\begin{equation}
\hat{V}^{\mathrm{out},\mathrm{in}}\left(  t\right)  =R\hat{I}^{\mathrm{out}%
,\mathrm{in}}\left(  t\right)  ,
\end{equation}
we obtain
\begin{align}
\hat{I}\left(  t\right)   &  =\frac{1}{R}\hat{V}\left(  t\right)  -2\hat
{I}^{\mathrm{in}}\left(  t\right)  ,\\
&  =\frac{1}{R}\hat{V}\left(  t\right)  -\frac{2}{\sqrt{R}}\hat{A}%
^{\mathrm{in}}\left(  t\right)  .
\end{align}
For a dissipationless circuit with Hamiltonian $H_{bare}\left(  \hat{\Phi
},\hat{Q}\right)  $, where $\hat{\Phi}$ is the generalized flux of the node
electrically connected to the transmission line, and $\hat{Q}$ its canonically
conjugate operator (top panel of Fig. \ref{Nyquist-model}), we can write the
Langevin equation,
\begin{align}
\frac{\mathrm{d}}{\mathrm{d}t}\hat{Q}  &  =\frac{i}{\hbar}\left[
H_{bare},\hat{Q}\right]  -\hat{I},\nonumber\\
&  =\frac{i}{\hbar}\left[  H_{bare},\hat{Q}\right]  -\frac{\mathrm{d}%
}{R\mathrm{d}t}\hat{\Phi}+\frac{2}{\sqrt{R}}\hat{A}^{\mathrm{in}}\left(
t\right)  . \label{Langevin-example}%
\end{align}
The latter equation is just a particular case of the more general quantum
Langevin equation giving the time evolution of any operator $\hat{Y}$ of a
system with Hamiltonian $H_{bare}$, which is coupled to the semi-infinite
transmission line by an Hamiltonian term proportional to another system
operator $\hat{X}$,
\begin{align}
\frac{\mathrm{d}}{\mathrm{d}t}\hat{Y}  &  =\frac{i}{\hbar}\left[
H_{bare},\hat{Y}\right] \nonumber\\
&  +\frac{1}{2i\hbar}\left\{  \left[  \hat{X},\hat{Y}\right]  ,2R^{\zeta
/2}\hat{A}^{\mathrm{in}}\left(  t\right)  -R^{\zeta}\frac{\mathrm{d}%
}{\mathrm{d}t}\hat{X}\right\}  .\nonumber\\
&  \label{general-QLE1}%
\end{align}
The value of $\zeta$ in Eq. (\ref{general-QLE1}) depends on whether the
damping is \textquotedblleft parallel" ($\zeta=-1$) or \textquotedblleft
series" type ($\zeta=+1$) [see Fig. \ref{Nyquist-model}]. In the parallel
case, the greater the line impedance the smaller the damping, whereas in the
series case the situation is reversed.

Equation (\ref{general-QLE1}) should be supplemented by
\begin{equation}
\left[  \hat{A}^{\mathrm{in}}\left(  t_{1}\right)  ,\hat{A}^{\mathrm{in}%
}\left(  t_{2}\right)  \right]  =\frac{i\hbar}{2}\frac{\partial}%
{\partial\left(  t_{1}-t_{2}\right)  }\delta\left(  t_{1}-t_{2}\right)
\end{equation}
and
\begin{equation}
\hat{A}^{\mathrm{out}}\left(  t\right)  =\zeta\left[  \hat{A}^{\mathrm{in}%
}\left(  t\right)  -R^{\zeta/2}\frac{\mathrm{d}}{\mathrm{d}t}\hat{X}\right]  .
\end{equation}
It follows from the last three equations that the output fields have the same
commutation relation as the input fields
\begin{equation}
\left[  \hat{A}^{\mathrm{out}}\left(  t_{1}\right)  ,\hat{A}^{\mathrm{out}%
}\left(  t_{2}\right)  \right]  =\frac{i\hbar}{2}\frac{\partial}%
{\partial\left(  t_{1}-t_{2}\right)  }\delta\left(  t_{1}-t_{2}\right)  .
\end{equation}
%
%
\subsection*{Quantum Langevin equation in the RWA approximation}
%
%
We now consider an approximate form of the input-output formalism which is
valid when the system degree of freedom consists of an oscillator with very
low damping, and for which all the frequencies of interest will lie in a
narrow range around the oscillator frequency $\omega_{a}$. We start from Eq.
(\ref{Langevin-example}) and use
\begin{align}
\hat{\Phi}  &  =\Phi^{ZPF}\left(  a+a^{\dag}\right)  ,\\
\hat{Q}  &  =Q^{ZPF}\frac{\left(  a-a^{\dag}\right)  }{i},
\end{align}
where $\Phi^{ZPF}=\sqrt{\hbar Z_{a}/2}$ and $Q^{ZPF}=\sqrt{\hbar/2Z_{a}}$.

We then obtain, neglecting the effect of driving terms oscillating at twice
the resonance frequency,
\begin{equation}
\frac{\mathrm{d}}{\mathrm{d}t}a=\frac{i}{\hbar}\left[  H_{bare},a\right]
-\omega_{a}\frac{Z_{a}}{2R}a+\sqrt{\frac{2Z_{a}}{\hbar R}}\tilde
{A}^{\mathrm{in}}\left(  t\right)
\end{equation}
with
\begin{equation}
\tilde{A}^{\mathrm{in}}(t)=\int_{0}^{\infty}\hat{A}^{\mathrm{in}}%
[\omega]e^{-i\omega t}\mathrm{d}\omega.
\end{equation}
The field amplitude $\tilde{A}^{\mathrm{in}}(t)$ is non-hermitian and contains
only the negative frequency component of $A^{\mathrm{in}}(t)$. For signals in
a narrow band of frequencies around the resonance frequency, we can make the
substitution
\begin{equation}
\sqrt{\frac{2}{\hbar\omega_{a}}}\tilde{A}^{\mathrm{in}}\left(  t\right)
\rightarrow\tilde{a}^{\mathrm{in}}\left(  t\right)  ,
\end{equation}
where
\begin{equation}
\tilde{a}^{\mathrm{in}}(t)=\int_{0}^{\infty}a^{\mathrm{in}}[\omega]e^{-i\omega
t}\mathrm{d}\omega.
\end{equation}
The input field operator $a^{\mathrm{in}}[\omega]$ is identical to
$a^{\leftarrow}[\omega]$ of the infinite line. We finally arrive at the RWA
quantum Langevin equation, also referred to in the quantum optics literature
as the quantum Langevin equation in the Markov approximation
\begin{equation}
\frac{\mathrm{d}}{\mathrm{d}t}a=\frac{i}{\hbar}\left[  H_{bare},a\right]
-\frac{\gamma_{a}}{2}a+\sqrt{\gamma_{a}}\tilde{a}^{\mathrm{in}}\left(
t\right)  ,
\end{equation}
where
\begin{equation}
\left[  \tilde{a}^{\mathrm{in}}\left(  t\right)  ,\tilde{a}^{\mathrm{in}%
}\left(  t^{\prime}\right)  ^{\dagger}\right]  =\delta\left(  t-t^{\prime
}\right)  .
\end{equation}
For any oscillator, the input output relationship is obtained from
\begin{equation}
\sqrt{\gamma_{a}}a\left(  t\right)  =\tilde{a}^{\mathrm{in}}\left(  t\right)
-\zeta\tilde{a}^{\mathrm{out}}\left(  t\right)  . \label{IOT}%
\end{equation}
It is worth noting that although $a^{\mathrm{in}}$ and $a^{\mathrm{out}}$ play
the role of $a^{\leftarrow}$ and $a^{\rightarrow}$ in Eq. (\ref{bosonic_com}),
only the average values of the moments of $a^{\mathrm{in}}$ can be imposed,
$a^{\mathrm{out}}$ being a \textquotedblleft slave" of the dynamics of
$a^{\mathrm{in}}$, as processed by the oscillator.
%
%

\begin{thebibliography}{99}
\bibitem {SumofEMenergy}Its value can also be obtained as the sum of the
electrostatic and magnetic energy in the line
\[
\frac{1}{2}\int_{-\infty}^{+\infty}\left\{  C_{\ell}\left[  V\left(
x,t\right)  \right]  ^{2}+L_{\ell}\left[  I\left(  x,t\right)  \right]
^{2}\right\}  dx.
\]
\end{thebibliography}

%
%
%
%
\end{document}